\documentclass[prd,preprint,tightenlines,floatfix,showpacs,
preprintnumbers,nofootinbib,eqsecnum]{revtex4-1}

\usepackage[T1]{fontenc}		% fnot encoding with polish letters !or OT4

\usepackage{amsmath,amsfonts,amssymb,amstext,mathrsfs}
\usepackage{mathpazo}

\usepackage[dvips]{graphicx}
\usepackage{epsf,float}
\usepackage{revsymb}

\usepackage{dcolumn}% Align table columns on decimal point
\usepackage{braket}
\usepackage{color,xcolor}
\usepackage{graphicx}
\usepackage{subfigure}
\usepackage{multirow}
\usepackage{tabularx}
\usepackage{pstricks}
\usepackage[section]{placeins}
\usepackage{booktabs}
\usepackage{array}

\usepackage{hyperref}

\def\Pom{I\!\!P}
\def\Reg{I\!\!R}

\begin{document}

\title{Light-by-light scattering in ultraperipheral $PbPb$ collisions \\
at the Large Hadron Collider}

\author{Mariola K{\l}usek-Gawenda}
\email{Mariola.Klusek@ifj.edu.pl}
\author{Piotr Lebiedowicz}
\email{Piotr.Lebiedowicz@ifj.edu.pl}
%\affiliation{Institute of Nuclear Physics PAN, PL-31-342 Krak\'ow, Poland}
\author{Antoni Szczurek\footnote{Also at University of Rzesz\'ow, PL-35-959 Rzesz\'ow, Poland.}}
\email{Antoni.Szczurek@ifj.edu.pl}
\affiliation{Institute of Nuclear Physics, Polish Academy of Sciences, Radzikowskiego 152,
PL-31-342 Krak\'ow, Poland}

\date{\today}

\vspace{2cm}

\begin{abstract}
We calculate cross sections for diphoton production
in (semi)exclusive $PbPb$ collisions, relevant for the LHC.
The calculation is based on equivalent photon approximation in
the impact parameter space.
The cross sections for elementary $\gamma \gamma \to \gamma \gamma$ 
subprocess are calculated including two different mechanisms.
We take into account box diagrams 
with leptons and quarks in the loops. 
In addition, we consider a vector-meson dominance (VDM-Regge) contribution
with virtual intermediate hadronic (vector-like) excitations of the photons.
We get much higher cross sections in $PbPb$ collisions 
than in earlier calculation from the literature.
This opens a possibility to study the $\gamma \gamma \to \gamma \gamma$
(quasi)elastic scattering at the LHC.
We present many interesting differential distributions which could
be measured by the ALICE, CMS or ATLAS Collaborations at the LHC.
We study whether a separation or identification of different components
(boxes, VDM-Regge) is possible. We find that the cross section for elastic
$\gamma \gamma$ scattering could be measured in the heavy-ion collisions 
for subprocess energies smaller than $W_{\gamma\gamma} \approx 15-20$~GeV.
\end{abstract}

\pacs{	25.75.Cj,
    	 	25.70.Bc,
    	 	34.50.-s }

%\pacs{12.60.Fr,13.85.-t,14.80.Da}

\maketitle

%----------------------------
\section{Introduction}
%----------------------------

In classical Maxwell theory photons/waves/wave packets do not interact.
In contrast, in quantal theory they can interact via quantal fluctuations.
So far only inelastic processes, i.e. production of hadrons or jets
via photon-photon fusion could be measured e.g. in $e^+ e^-$ collisions
\cite{Buskulic:1993mm,Muramatsu:1994rq,Ackerstaff:1996ms,Barate:1999qx}\footnote{Please note that here the incoming photons are virtual.}.

The light-by-light scattering to the leading and next-to-leading
order was discussed earlier in the literature, see \cite{Bohm:1994sf,Jikia:1993tc,Bern:2001dg}
also in the context of search for effects of new particles
and interactions, e.g. see \cite{Gounaris:1998qk,Gounaris:1999gh}.
The cross section for elastic $\gamma \gamma \to \gamma \gamma$ 
scattering is so small that till recently it was beyond 
the experimental reach.
In $e^+ e^-$ collisions the energies and/or couplings of photons to electrons/positrons
are rather small so that the corresponding $\gamma\gamma \to \gamma\gamma$ 
cross section is extremely small.
A proposal to study helicity dependent $\gamma\gamma \to \gamma\gamma$
scattering in the region of MeV energies with the help of high
power lasers was discussed recently e.g. in Ref.~\cite{Homma:2015fva}.

In proton-proton collisions the subprocess energies
(diphoton invariant masses) can be larger and 
the underlying photon-photon scattering is possible 
in exclusive processes \cite{d'Enterria:2013yra,Lebiedowicz:2013fta,Fichet:2013gsa}.
However, at low two-photon invariant masses there is a competitive
diffractive QCD mechanism through the $gg \to \gamma\gamma$ subprocess
with quark boxes \cite{Khoze:2004ak,Lebiedowicz:2012gg,Harland-Lang:2014lxa} 
which gives much higher cross section than the photon-photon fusion 
\cite{Lebiedowicz:2013fta}. 
The reader may find a detailed comparison of the two mechanisms in chapter 5 
of \cite{PhD:Lebiedowicz}. 
The QCD mechanism provides an explanation
of experimental cross sections measured by the CDF Collaboration
\cite{Aaltonen:2007am,Aaltonen:2011hi}.

Ultraperipheral collisions (UPC) of heavy-ions provide a nice
possibility to study several two-photon induced processes such as:
$\gamma\gamma \to l^+l^-$, $\gamma\gamma \to \pi^+\pi^-$,
$\gamma\gamma \to $ dijets, etc. (see e.g.~\cite{Baur:2001jj,Bertulani:2005ru,Baltz:2007kq}).
It was realized only recently that ultraperipheral heavy-ions 
collisions can be also a good place where photon-photon elastic
scattering could be tested experimentally \cite{d'Enterria:2013yra}.
In Ref.~\cite{d'Enterria:2013yra} a first estimate of the corresponding 
cross section was presented.

In this paper we present a more detailed study
with more realistic approach and show
several differential distributions not discussed so far.
We include also a new, higher order, 
mechanism not discussed so far in the literature. 

%----------------------------------------------------------------------------
\section{$\gamma \gamma \to \gamma \gamma$ elementary cross section}
%----------------------------------------------------------------------------

Before presenting the nuclear cross sections let us concentrate
first on elementary $\gamma \gamma \to \gamma \gamma$ scattering.

\begin{figure}[!h]
\includegraphics[scale=0.4]{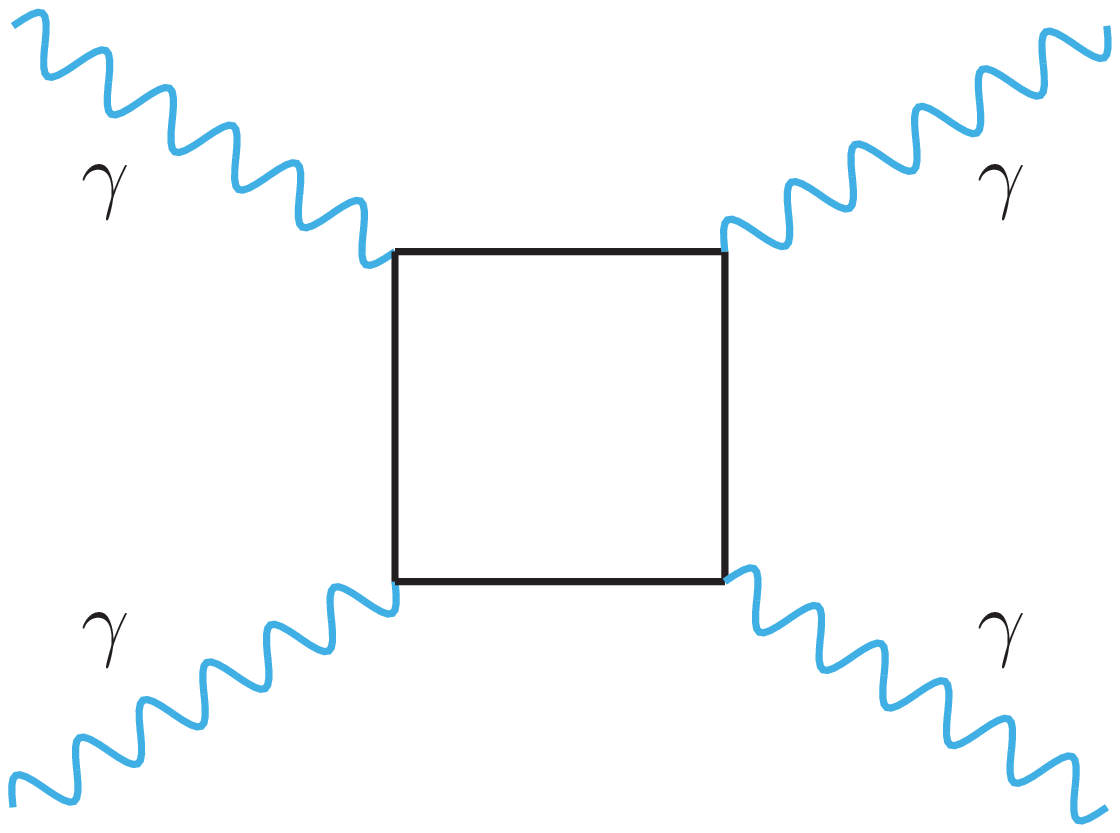}
\includegraphics[scale=0.4]{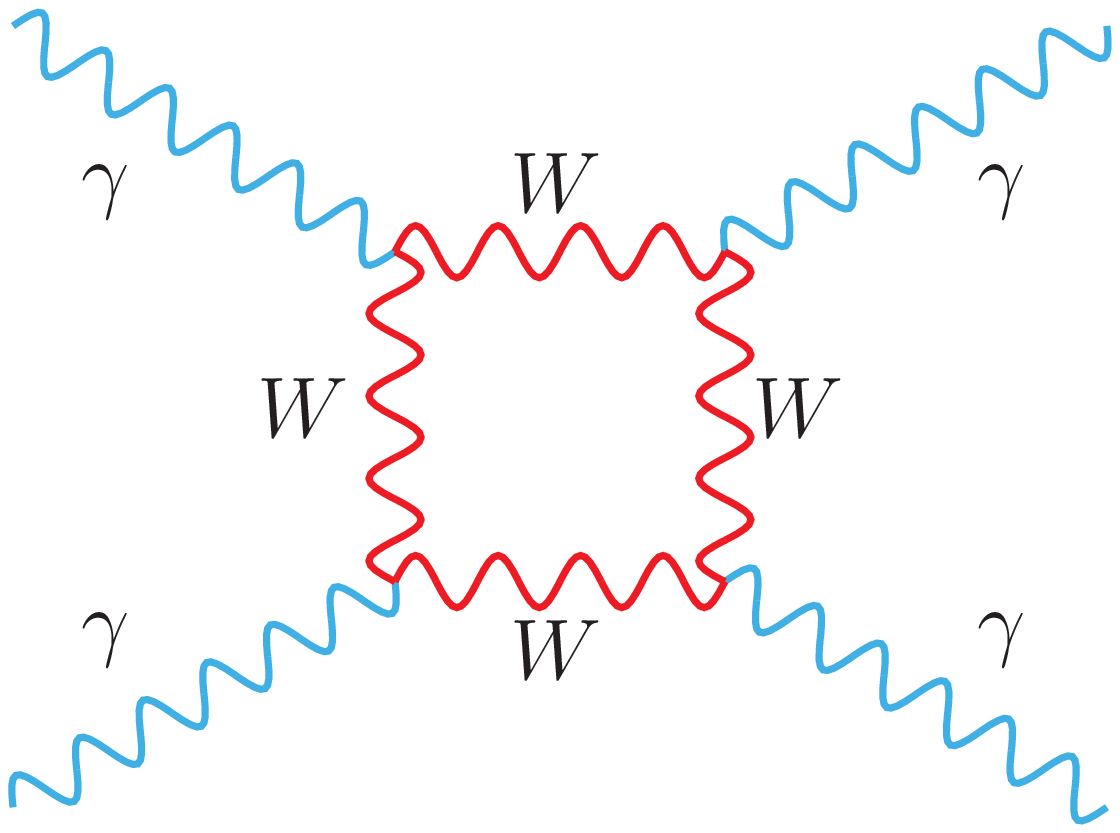}
\caption{Light-by-light scattering mechanisms with 
the lepton and quark loops (left panel) and as an example one topology of diagram
for intermediate the $W$-boson loop (right panel).}
\label{fig:diagrams_boxes}
\end{figure}

The lowest order QED mechanisms with elementary particles are shown
in Fig.~\ref{fig:diagrams_boxes}. The diagram in the left panel
is for lepton and quark (elementary fermion) loops, while the diagram
in the right panel is for $W$ (spin-1) boson loops. 
The mechanism on the left hand side dominates
at lower photon-photon energies while the mechanism on the right hand side becomes 
dominant at higher photon-photon energies 
(see e.g.~\cite{Bardin:2009gq,Lebiedowicz:2013fta}).
In numerical calculations here we include box diagrams with fermions only,
which will be explained in the following.

%-----------------------------------------------------------------------------
%-----------------------------------------------------------------------------

The one-loop box diagrams were calculated by using
the Mathematica package {\tt{FormCalc}} \cite{Hahn:1998yk} and the {\tt{LoopTools}}
library based on \cite{vanOldenborgh:1989wn} to evaluate one-loop integrals.
The complete matrix element was generated in terms of 
two-, three- and four-point coefficient functions \cite{Passarino:1978jh},
internally-defined photon polarisation vectors,
and kinematic variables (four-momenta of incoming and outgoing photons).
Our result was confronted with that in 
\cite{Jikia:1993tc,Bern:2001dg,Bardin:2009gq}.

\begin{figure}[!h]
\includegraphics[scale=0.5]{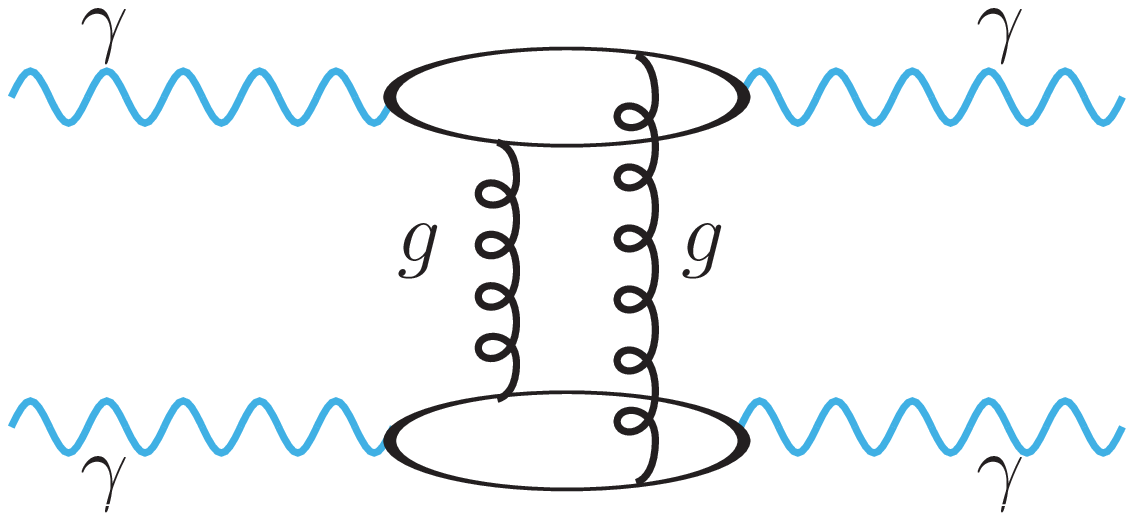}
\includegraphics[scale=0.5]{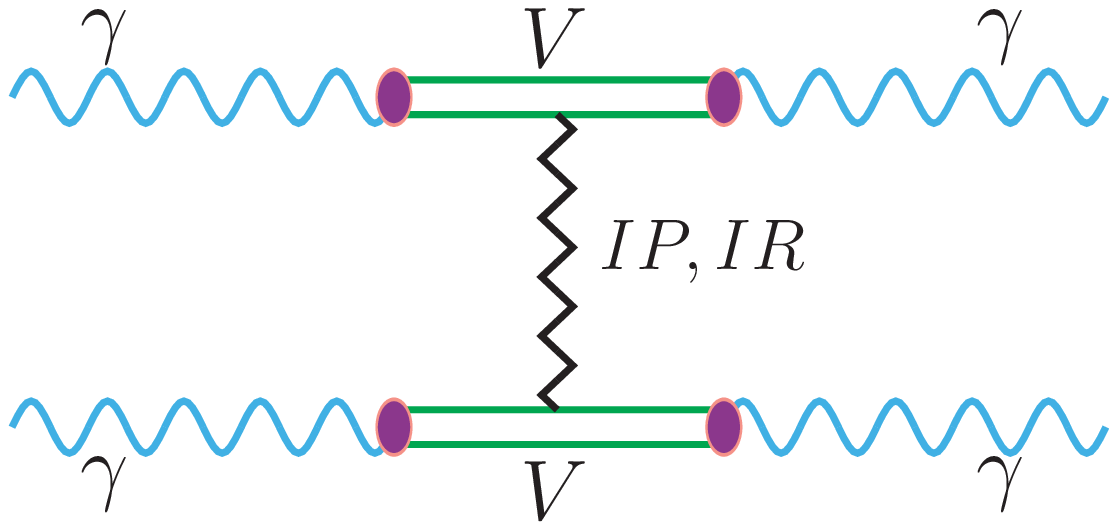}
\caption{Other elementary $\gamma\gamma \to \gamma\gamma$ processes. 
The left panel represents two-gluon exchange and the right panel 
is for VDM-Regge mechanism.}
\label{fig:diagrams_t_channel}
\end{figure}

In principle, high-order contributions, not considered so far
in the context of elastic scattering, are possible too.
In Ref.~\citep{Bern:2001dg} the authors considered both 
the QCD and QED corrections (two-loop Feynman diagrams)
to the one-loop fermionic contributions in the ultrarelativistic limit
($\hat{s},|\hat{t}|,|\hat{u}| \gg m_f^2$). 
The corrections are quite small numerically,
showing that the leading order computations considered by us are satisfactory.
In Fig.~\ref{fig:diagrams_t_channel} (left panel) we show a process 
which is the same order in $\alpha_{em}$ but higher order in $\alpha_s$. 
This mechanism is formally three-loop type and therefore difficult for calculation. 
We will not consider here the contribution of this three-loop mechanism. 
The exact three-loop calculation for this process is not yet available.
Instead we shall consider ''a similar'' process shown 
in the right panel where both photons fluctuate into virtual vector mesons
(three different light vector mesons are included). 
In this approach the interaction ''between photons'' 
happens when both photons are in their hadronic states. 

%In all cases we assume the interaction parameters to be the same as 
%for the $\pi^0 p$ interaction. Some parameters are also the same as for
%the VDM-Regge model for the $\gamma \gamma \to \rho^0 \rho^0$ reaction
%\cite{Klusek:2009yi}.

%-----------------------------------------------------------------------------
%-----------------------------------------------------------------------------

The differential cross section for the elementary 
$\gamma \gamma \to \gamma \gamma$ subprocess can be calculated as:
\begin{equation}
\frac{d\sigma_{\gamma\gamma \to \gamma\gamma}}{dt} 
= \frac{1}{16 \pi s^2} \overline{\left| \mathcal{A}_{\gamma\gamma \to
    \gamma\gamma} \right|^2} 
\end{equation}
or
\begin{equation}
\frac{d\sigma_{\gamma\gamma \to \gamma\gamma}}{d\Omega} 
= \frac{1}{64 \pi^2 s} \overline{\left| \mathcal{A}_{\gamma\gamma \to
    \gamma\gamma} \right|^2} \; .
\end{equation}

In the most general case, including virtualities of initial photons,
the amplitude can be written as:
$\mathcal{A} = \mathcal{A}_{TT} + \mathcal{A}_{TL}
             + \mathcal{A}_{LT} + \mathcal{A}_{LL}$ 
where $\mathcal{A}_{TL} \propto \sqrt{Q_2^2}$,
      $\mathcal{A}_{LT} \propto \sqrt{Q_1^2}$,
      $\mathcal{A}_{LL} \propto \sqrt{Q_1^2 Q_2^2}$.
Since in UPC's $Q_1^2, Q_2^2 \approx 0$ (nuclear form factors kill large virtualities)
the other terms can be safely neglected and $\mathcal{A} \approx \mathcal{A}_{TT}$.

The amplitude for the VDM-Regge contribution 
(see Fig.~\ref{fig:diagrams_t_channel})
can be written as\footnote{In fact the helicity amplitude 
can be written as
$A_{\lambda_1 \lambda_2 \to \lambda_3 \lambda_4} = 
\delta_{\lambda_1 \lambda_3} \delta_{\lambda_2 \lambda_4} \cdot A$}
\begin{equation}
\mathcal{A}_{\gamma\gamma \to \gamma\gamma}(s,t) \approx
\left( \sum\limits_{i=1}^{3} C^2_{\gamma \to V_i} \right) 
 \mathcal{A}\left(s,t\right) \exp\left( \frac{B}{2} t \right)
\left( \sum\limits_{j=1}^{3} C^2_{\gamma \to V_j} \right) \; ,
\label{eq:A_gg_gg_st}
\end{equation}
where $i,j=\rho, \omega, \phi$.
The $\gamma \to V$  transition constants are taken from 
\cite{Ioffe:1985ep} (see chapter 5, Eq. (1.11)). 
The amplitude for
$V_i V_j \to V_i V_j$ elastic scattering is parametrized in 
the Regge approach similar as for $\gamma\gamma \to \rho^0\rho^0$ 
in Ref.~\cite{Klusek:2009yi}
\begin{equation}
 \mathcal{A}\left(s,t\right)  \approx s
   \left( \left(1+i \right) 
C_{\Reg} \left(\frac{s}{s_0}\right)^{\alpha_{\Reg} \left(t\right)-1}
+ i 
C_{\Pom} \left( \frac{s}{s_0} \right)^{\alpha_{\Pom} \left(t\right)-1}
\right) \; .
\end{equation}

In all cases we assume the interaction parameters to be the same as for the 
$\pi^0 p$ interaction and obtained by the averaging:
\begin{equation}
\mathcal{A}_{\pi^0 p}(s,t) = \frac{1}{2} \left( \mathcal{A}_{\pi^+ p}(s,t) 
                                    + \mathcal{A}_{\pi^- p}(s,t) \right) \; .
\label{pi0p_amplitude}
\end{equation}
Our amplitude here are normalized such that the optical theorem reads 
(for massless particles):
\begin{equation}
\sigma_{\pi p}^{tot}(s) = \frac{1}{s} Im \mathcal{A}_{\pi p}(s,t=0) \; .
\label{sigma_tot_amplitude}
\end{equation}
Some parameters ($C_{\gamma \to \rho^0}$, 
$C_{\Reg}$, $C_{\Pom}$) are also the same as for
the VDM-Regge model for $\gamma \gamma \to \rho^0 \rho^0$
\cite{Klusek:2009yi}.
The parameters ($C_{\Reg}$ and $C_{\Pom}$) 
are fixed assuming Regge factorization and the Donnachie-Landshoff 
parametrizations  \cite{Donnachie:1992ny}
of the $NN$ and $\pi N$ total cross sections
(see e.g.~\cite{Donnachie:2002en,Szczurek:2001py}).
The slope parameter (see Eq.~(\ref{eq:A_gg_gg_st})), 
in general a free parameter, should be similar as for 
the pion-pion (dipole-dipole) scattering. 
For a simple estimate here we take $B =4$~GeV$^{-2}$
as in our previous paper on double $\rho^0$ production \cite{Klusek:2009yi}.

%-----------------------------------------------------------------------------
\begin{figure}[!h]
\includegraphics[scale=0.35]{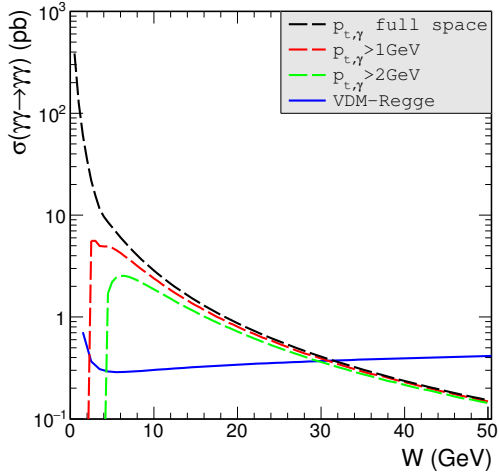}
\caption{Integrated $\gamma\gamma \to \gamma\gamma $ cross section
as a function of the subsystem energy. 
The dashed lines show contribution of boxes and the solid line represents result
of the VDM-Regge mechanism.}
\label{fig:sig_elem}
\end{figure}
%----------------------------------------------------------------------------

The elementary angle-integrated cross section for the box and 
VDM-Regge contributions is shown in Fig.~\ref{fig:sig_elem}
as a function of the photon-photon subsystem energy.
Lepton and quark amplitudes interfere in cross section
for the box contribution.
For instance in the $4<W<50$ GeV region, 
neglecting interference effects, the lepton contribution 
to the box cross section is by a factor $5$ bigger than the quark contribution.
Interference effects are, however, large and cannot be neglected.
At energies $W >30$~GeV the VDM-Regge cross section becomes larger
than that for the box diagrams. Can this be seen/identified 
in heavy ion lead-lead collisions at the LHC including experimental cuts? 
We will try to answer this question in this paper.

\begin{figure}[!h]
\includegraphics[scale=0.4]{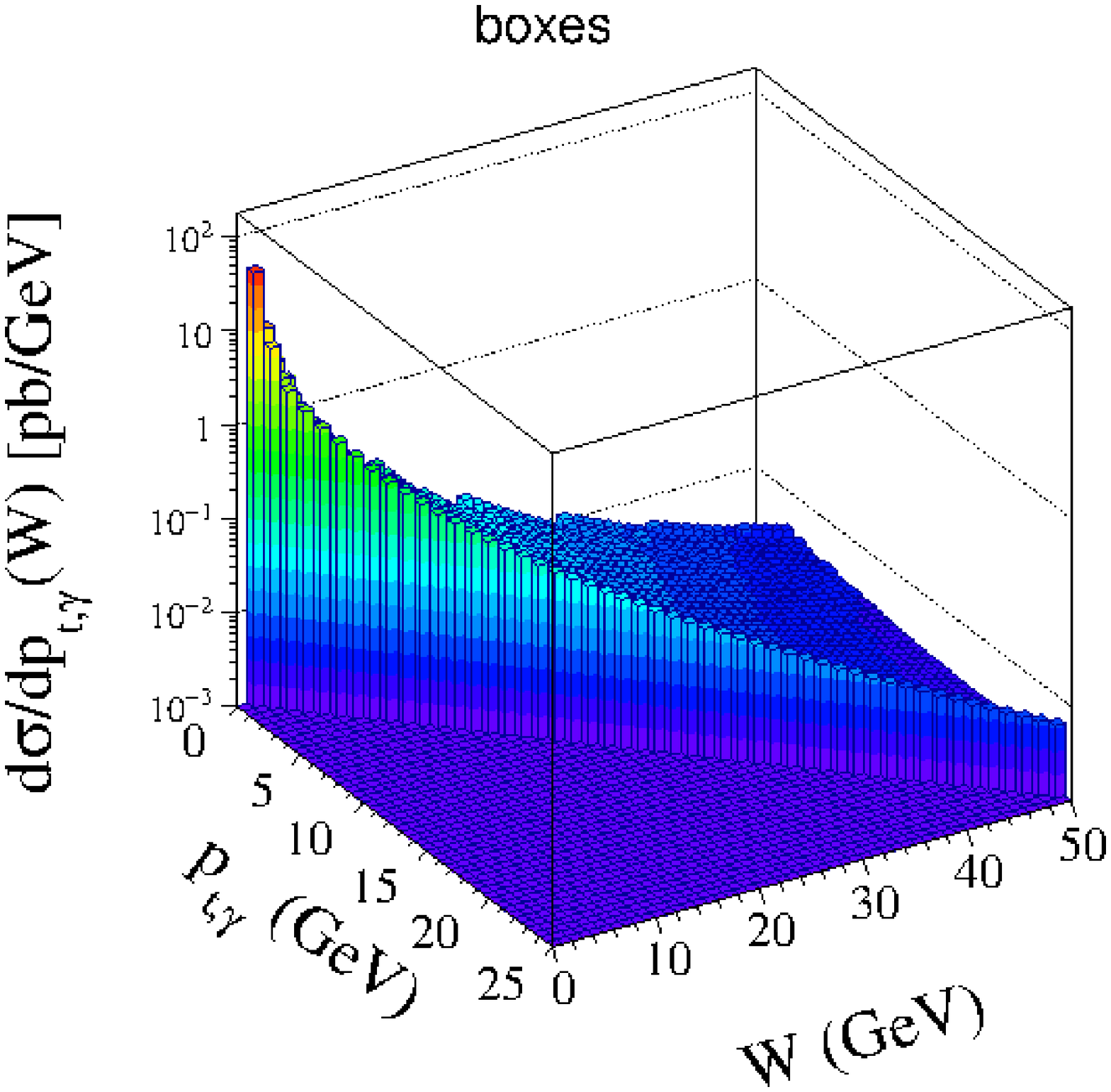}
\includegraphics[scale=0.4]{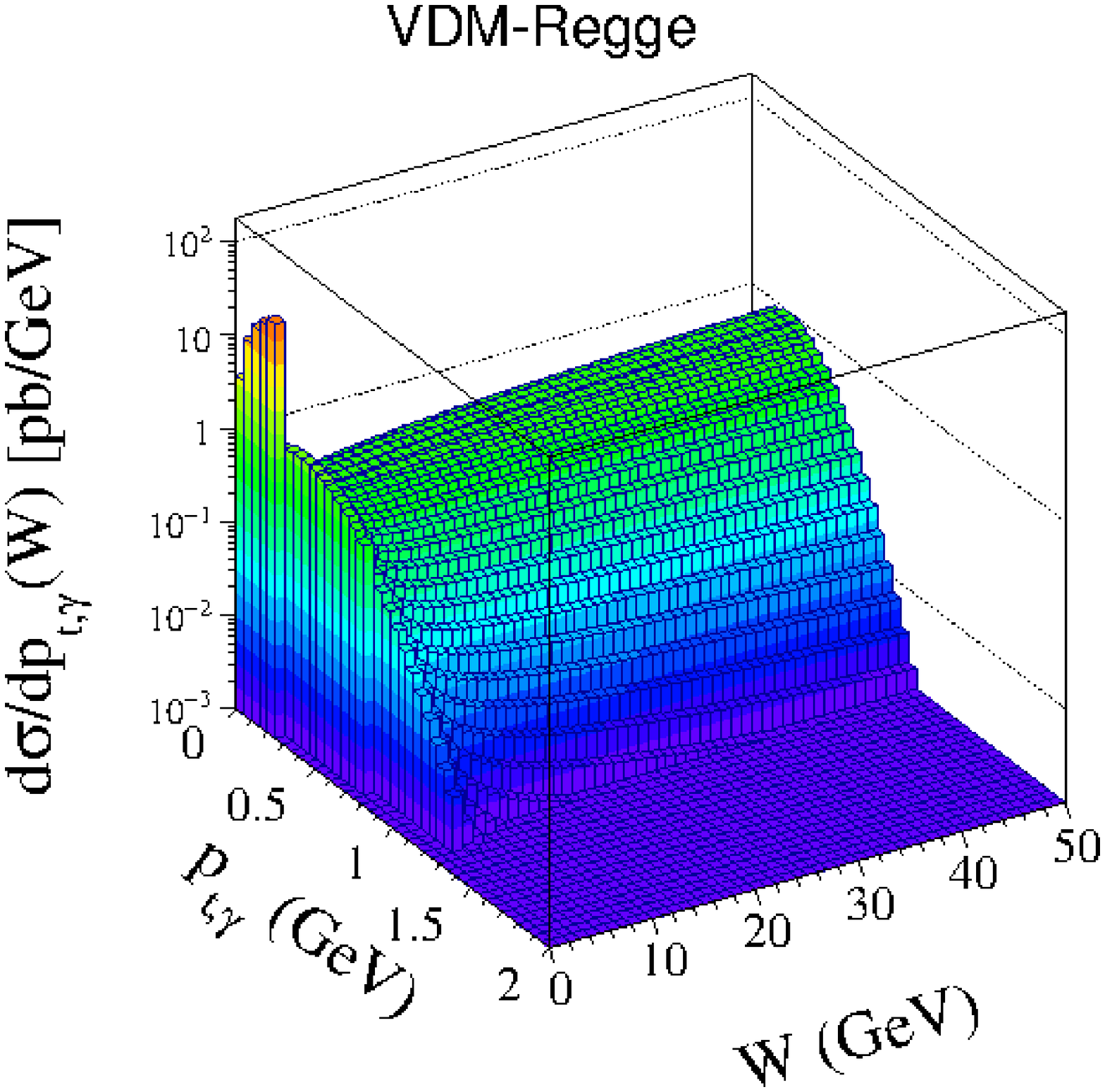}
\caption{Elementary cross section $d \sigma/dp_{t,\gamma}$ 
as a function of the subprocess energy $W$  
($\gamma\gamma$ invariant mass in the nuclear process) 
and transverse momentum of one of the
outgoing photons for the box (left panel) and VDM-Regge (right panel)
mechanisms.}
\label{fig:dsig_dpt_W}
\end{figure}

For completeness in Fig.~\ref{fig:dsig_dpt_W} we show also differential 
cross section for the box (left panel) and the VDM-Regge (right panel) 
components as a function of subsystem energy and photon 
transverse momentum.
The distribution for the box mechanism (left panel) has a characteristic
enhancement for $p_{t,\gamma} \approx W/2$ which is due to jacobian
of variables transformation from finite $d\sigma /dz$ distribution at $z=0$. 
One can observe a fast fall-off of the differential cross section 
with photon transverse momenta for the VDM-Regge mechanism (right panel). 
Imposing lower $p_{t,\gamma}$ cuts in experiments would therefore almost 
completely kill the VDM-Regge contribution.
We expect that compared to our soft VDM-Regge component 
the two-gluon exchange component (see the left panel of 
Fig.~\ref{fig:diagrams_t_channel}) should have larger wings/tails 
at larger transverse momenta. This may be a bit academic problem
but may be interesting by itself.

%-----------------------------------------------------------------------
%-----------------------------------------------------------------------

\begin{figure}[!h]
\includegraphics[scale=0.4]{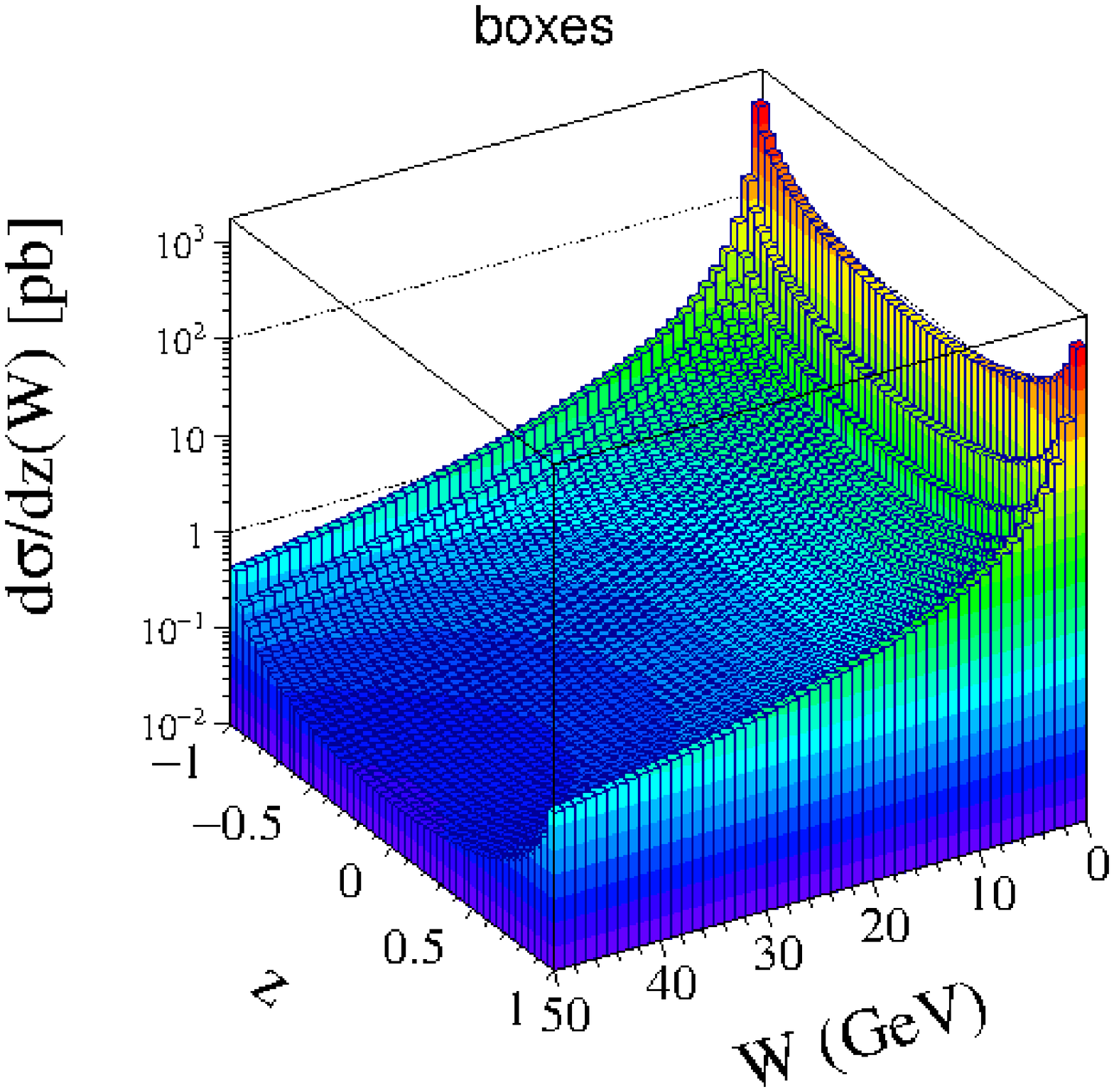}
\includegraphics[scale=0.4]{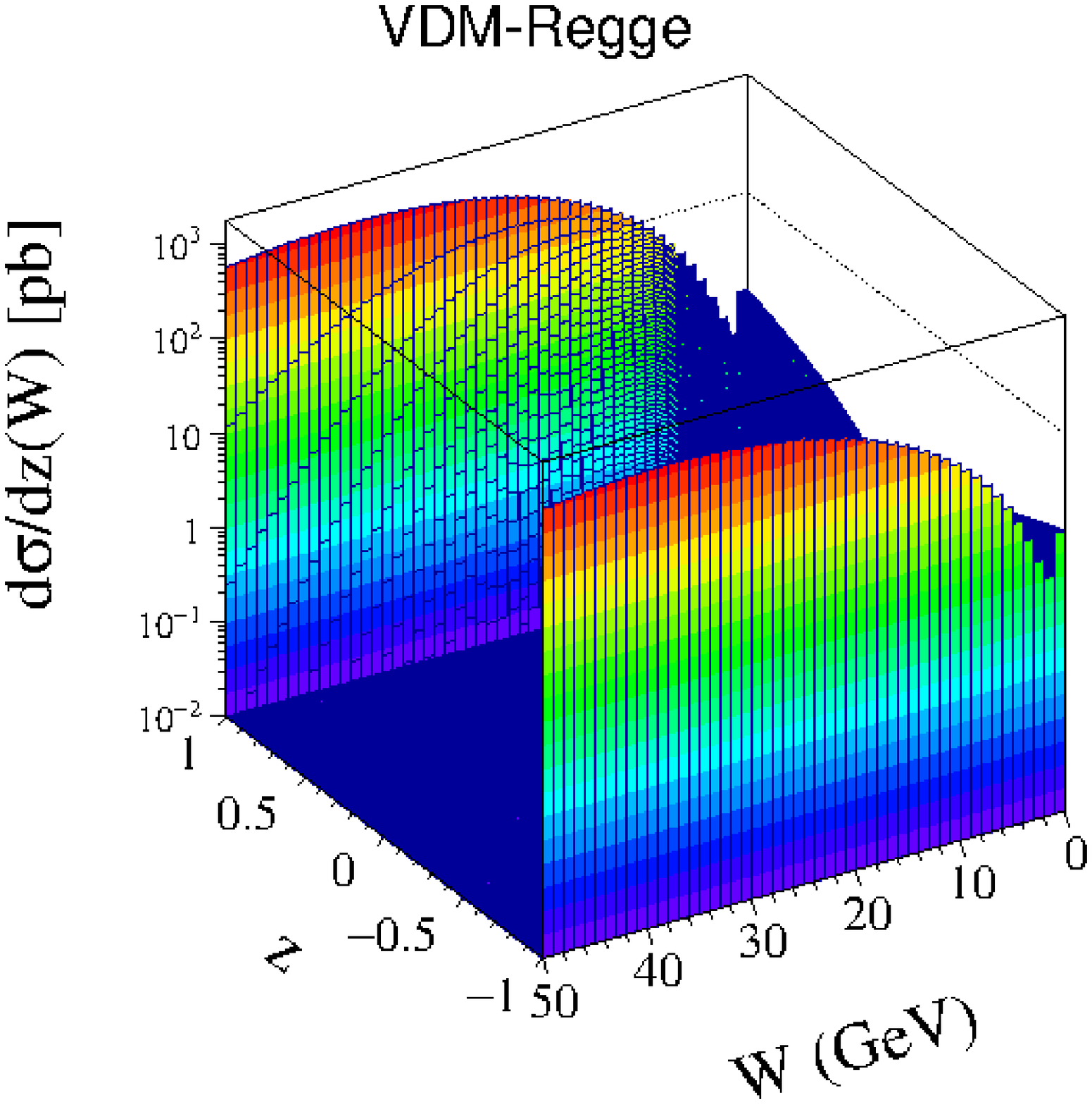}
\caption{Elementary cross section $d \sigma/dz$ 
as a function of the subprocess energy $W$ 
($\gamma\gamma$ invariant mass in the nuclear process) and $z=\cos \theta$ 
for the box (left panel) and VDM-Regge (right panel)
mechanisms.}
\label{fig:dsig_dz_W}
\end{figure}

Fig.~\ref{fig:dsig_dz_W} presents two-dimensional distribution
of the elementary $\gamma\gamma \to \gamma \gamma$ cross section
as a function of cosine of the angle between outgoing photons $z=\cos \theta$
and energy. The left panel shows distribution 
for the box mechanism and the right panel is for the VDM-Regge mechanism.
For both cases, the largest cross section occurs at $z \approx \pm 1$.

%-----------------------------------------------------------------------
%-----------------------------------------------------------------------

Now we shall proceed to nuclear calculations where the elementary cross
sections discussed above are main ingredients of the approach.

%----------------------------------------
\section{Diphoton production in UPC of heavy ions}
%----------------------------------------

%
%---------------------------------------------------------------------------
\begin{figure}[!h]
\includegraphics[scale=0.4]{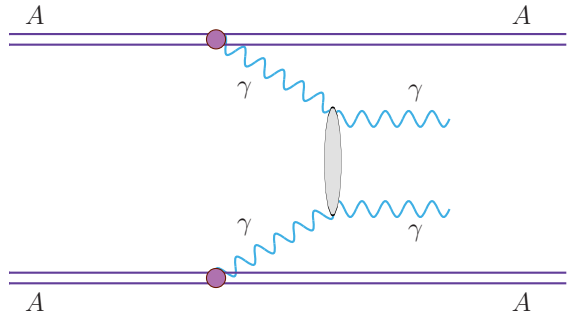}
\caption{$AA \to AA \gamma \gamma$ in ultrarelativistic UPC of heavy
  ions.}
\label{fig:diagram_AA_AAgamgam}
\end{figure}
%----------------------------------------------------------------------------
%
The general situation for $AA \to AA \gamma \gamma$ 
is sketched in Fig.~\ref{fig:diagram_AA_AAgamgam}.
Here we follow our earlier approach applied already to
different reactions \cite{Klusek:2009yi,KlusekGawenda:2010kx,
KlusekGawenda:2010jq,KlusekGawenda:2011ib,Baranov:2012vu,Klusek-Gawenda:2013rtu,
Klusek-Gawenda:2013dka}.
In our equivalent photon approximation in the impact parameter space, 
the total (phase space integrated) cross section 
is expressed through the five-fold integral
(for more details see e.g.~\cite{KlusekGawenda:2010kx})
\begin{eqnarray}
\sigma_{A_1 A_2 \to A_1 A_2 X}\left(\sqrt{s_{A_1A_2}} \right) &=&\int \sigma_{\gamma \gamma \to \gamma \gamma} 
\left(\sqrt{s_{A_1A_2}} \right)
N\left(\omega_1, {\bf b_1} \right)
N\left(\omega_2, {\bf b_2} \right) 
S_{abs}^2\left({\bf b}\right)\nonumber \\
& \times & 
2\pi b \mathrm{d} b \, \mathrm{d}\overline{b}_x \, \mathrm{d}\overline{b}_y \, 
\frac{W_{\gamma\gamma}}{2}
\mathrm{d} W_{\gamma\gamma} \, \mathrm{d} Y_{\gamma \gamma} \;,
\label{eq:EPA_sigma_final_5int}
\end{eqnarray}
where $N(\omega_i,{\bf b_i})$ are photon fluxes\footnote{Nuclear charge form factors are main ingredients of the photon flux\cite{KlusekGawenda:2010kx}.} and 
\begin{equation}
Y_{\gamma \gamma}=\frac{1}{2}\left( y_{\gamma_1} + y_{\gamma_2} \right)
\label{eq:def_Y}
\end{equation} 
is a rapidity of the outgoing $\gamma \gamma$ system. 
The invariant mass of the $\gamma\gamma$ system is defined as
\begin{equation}
W_{\gamma\gamma}=\sqrt{4\omega_1\omega_2} \;,
\end{equation} 
where $\omega_{1/2} = W_{\gamma\gamma}/2 \exp(\pm Y_{\gamma\gamma})$.
The quantities $\overline{b}_x$, $\overline{b}_y$ are the components of 
the ${\bf b_1}$ and ${\bf b_2}$ vectors:
\begin{equation}
{\bf b_1} = \left[ \overline{b}_x + \frac{b}{2}, \overline{b}_y \right] \;, \qquad
{\bf b_2} = \left[ \overline{b}_x - \frac{b}{2}, \overline{b}_y \right] \;.
\end{equation} 

Eq. \eqref{eq:EPA_sigma_final_5int} allows to calculate total cross section,
distributions in the impact parameter ($b=b_m$), 
invariant mass of the diphoton system ($W_{\gamma\gamma}=M_{\gamma \gamma}$) 
or rapidity of the pair ($Y_{\gamma \gamma}$) of these particles.
$S^2_{abs}(b)$ is a geometrical factor
which takes into account survival probability
of nuclei as a function of impact parameter. To a reasonable approximation
it can be approximated as
\begin{equation}
S^2_{abs} (b) = \theta \left( b- \left( R_A+R_B \right) \right) \;.
\end{equation}

If one wishes to impose some cuts on produced particles (photons) 
which come from experimental requirements or to have distribution 
in some helpful and interesting kinematical variables 
of individual particles (here photons), a more complicated calculations 
are required. 
Then we have to introduce into the integration an additional dimension 
related to angular distribution for the subprocess
(e.g. $z=\cos\theta$ or $p_{t,\gamma}$).
Then we define kinematical variables of photons in 
the $\gamma\gamma$ center-of-mass system (denoted here by $^*$):
\begin{eqnarray}
E^*_{\gamma_i} =	p^*_{\gamma_i}	 &=& \frac{W_{\gamma\gamma}}{2} \;, \\
z	= \cos\theta^*		&=& \sqrt{1-\left(\frac{p_{t,\gamma}}{p^*_{\gamma_i}}\right)^2} \;, \\
p_{z,\gamma_i}^*		&=& \pm z p^*_{\gamma_i} \;, \\
y_{\gamma_i}^* 	&=& \frac{1}{2} \ln 
\frac{E^*_{\gamma_i}+p_{z,\gamma_i}^*}{E^*_{\gamma_i}-p_{z,\gamma_i}^*} 
\end{eqnarray}
and in overall $AA$ center of mass system:
\begin{eqnarray}
y_{\gamma_i}		&=& Y_{\gamma\gamma} + y_{\gamma_i}^* \;, \\ 
p_{z,\gamma_i} 	&=& p_{t,\gamma} \sinh(y_{\gamma_i}) \;, \\
E_{\gamma_i}		&=& \sqrt{p_{z,\gamma_i}^2+p_{t,\gamma}^2} \;,
\end{eqnarray}
where $i=1,2$ means first or second outgoing photon, respectively.

%\newpage

%-----------------------------
\section{First results}
%-----------------------------

%-----------------------------------------------------------------------------------------------------------------------
\begin{table}[!h]
\begin{tabular}{|l|r|r|r|r|}
\hline
							&	\multicolumn{2}{c|}{boxes} 
							&   \multicolumn{2}{c|}{VDM-Regge} \\
cuts							&	$F_{realistic}$	& 	$F_{monopole}$	
							& 	$F_{realistic}$	& 	$F_{monopole}$ \\ \hline
$W_{\gamma\gamma}>5$ GeV		& 306	& 349	& 19		& 22 \\
$W_{\gamma\gamma}>5$ GeV, $p_{t,\gamma}>2$ GeV 
							& 159	& 182	& 7E-9	& 8E-9 \\
$E_{\gamma}>3$ GeV			& 16 692	& 18 400	& 13		& 14 \\
$E_{\gamma}>5$ GeV			& 4 800	& 5 450	& 4		& 6	\\
$E_{\gamma}>3$ GeV, $|y_{\gamma_i}|<2.5$
							& 183 	& 210 		& 7E-2	& 8E-2 \\
$E_{\gamma}>5$ GeV, $|y_{\gamma_i}|<2.5$		
							& 54		& 61		& 3E-4	& 6E-4	\\ 
$p_{t,\gamma}>0.9$ GeV, $|y_{\gamma_i}|<0.7$ (ALICE cuts)		
							& 107		& 		& 	& 	\\	 
$p_{t,\gamma}>5.5$ GeV, $|y_{\gamma_i}|<2.5$ (CMS cuts)		
							& 10		& 		& 	& 	\\					
\hline
\end{tabular}
\caption{ \small
Integrated cross sections in $nb$ for exclusive diphoton production 
processes with both photons measured for $\sqrt{s_{NN}}=5.5$ TeV (LHC).
The calculations was performed within impact-parameter EPA. 
The values of the total cross sections are shown for different cuts on
kinematic variables.
}
\label{table:cross_sections}
\end{table}
%-----------------------------------------------------------------------------------------------------------------------

To illustrate the general situation in Table~\ref{table:cross_sections}
we have collected integrated cross sections corresponding to different
kinematical cuts. Here we show results for the two (boxes, VDM-Regge) 
mechanisms separately\footnote{By doing so we neglect possible
interference effects between the two mechanisms.}
for very different kinematical situation. 
In all cases considered the cross section obtained with the monopole 
form factor is by more than $10 \%$ bigger than that obtained with the 
realistic form factor (Fourier transform of nucleus charge distribution).
In the first row we show results for cuts from Ref.~\cite{d'Enterria:2013yra}. 
We get much larger cross section than that found in \cite{d'Enterria:2013yra} 
($\sigma = 35 \pm 7$ nb).
In this case the VDM-Regge contribution not considered in earlier 
calculations constitutes about $7 \%$ of the dominant box contribution.
Already the cut on transverse momentum of photons as large as 
$p_{t,\gamma} >2$~GeV completely kills the VDM-Regge contribution which is very
forward/backward peaked. For the box contribution the effect is
much smaller.
The cut on photon-photon energies is not necessary. If we impose only 
cuts on energy of photons in the overall (nucleus-nucleus)
center-of-mass system (laboratory frame) the box-contribution 
is much larger, of the order of microbarns.
However, restricting to the rapidity coverage of the main detector 
diminishes the cross sections considerably, especially for the VDM-Regge 
contribution. The explanation will become clear when discussing 
differential distributions below.

\begin{figure}[!h]  
\center
\includegraphics[scale=0.35]{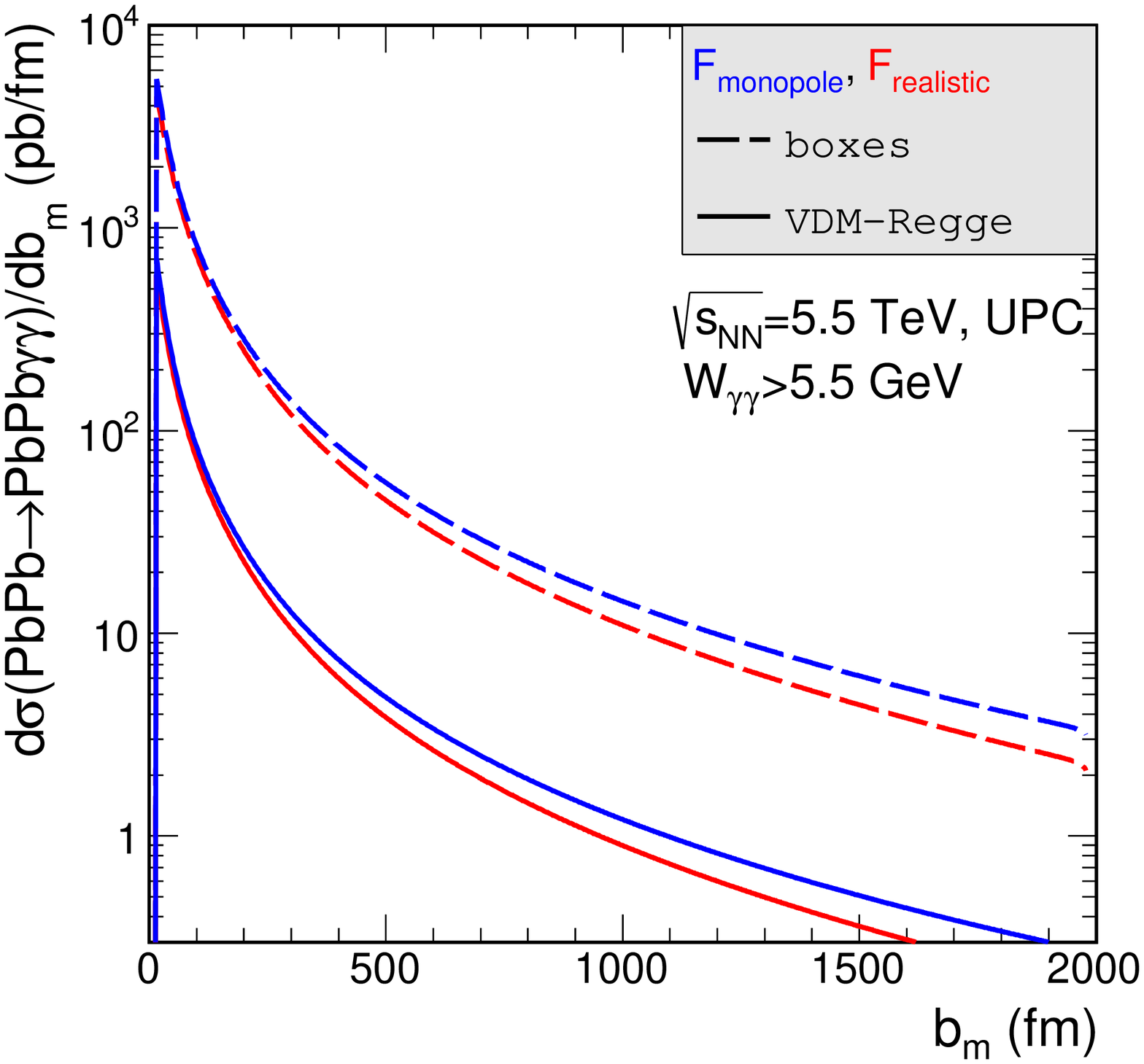}
\includegraphics[scale=0.35]{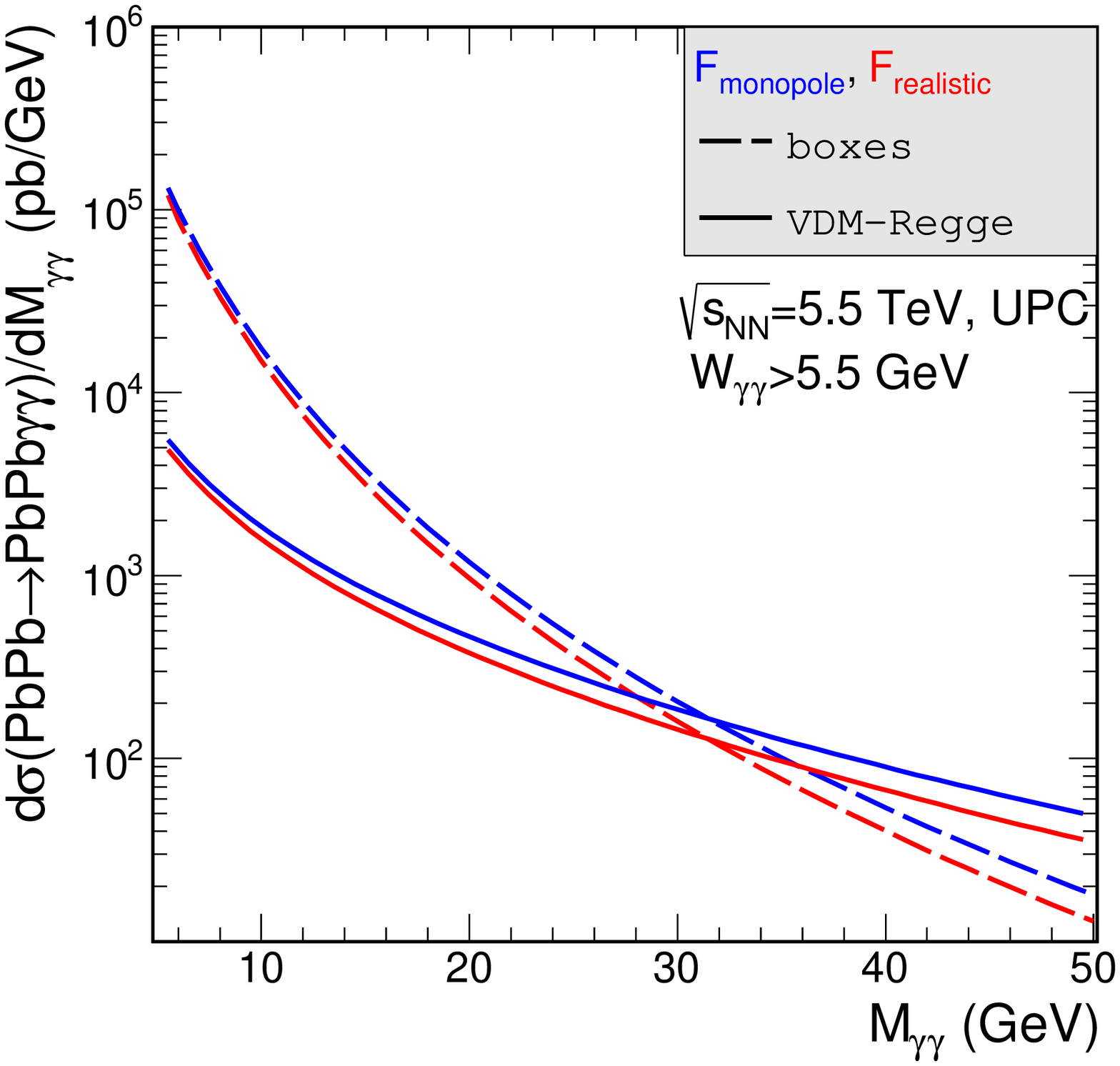}
\includegraphics[scale=0.35]{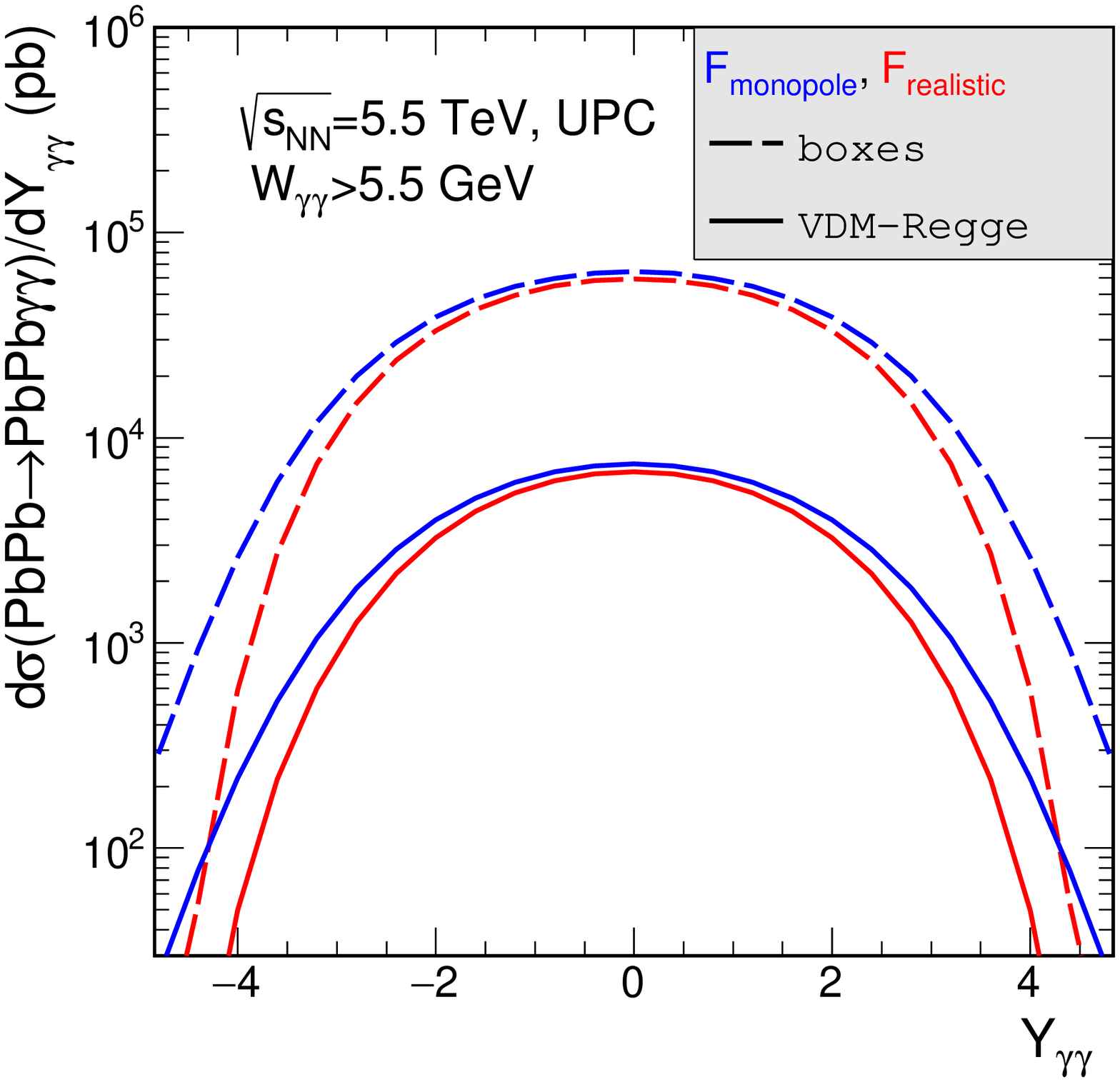}
  \caption{\label{fig:EPA_standard}
  \small
Predictions for the PbPb$\to$PbPb$\gamma\gamma$ reaction
in UPC of heavy ions.   
Differential nuclear cross section as a function of impact parameter,
$\gamma\gamma$ invariant mass and rapidity of photon pairs
at $\sqrt{s_{NN}}=5.5$ TeV and with extra cut on 
$W_{\gamma\gamma}>5.5$ GeV.
The distributions with realistic charge density are depicted by
the red (lower) lines and the distributions which are calculated
using monopole form factor are shown by the blue (upper) lines.
The dashed lines show the results for the case
when only box contributions (fermion loops) are included.
The solid lines show the results for the VDM-Regge mechanism.
}
\end{figure}

In Fig.~\ref{fig:EPA_standard} we show results which can
be obtained by calculating five-fold integral (see Eq.(\ref{eq:EPA_sigma_final_5int})).
In this calculation we have imposed only a lower cut (5.5 GeV) 
on photon-photon energy (or diphoton invariant mass) to get rid of 
the resonance region which may be more complicated.
Each of distributions (in $b_m$, $M_{\gamma\gamma}$, $Y_{\gamma\gamma}$ 
for boxes and VDM-Regge)
is shown for the case of realistic charge density 
and monopole form factor in nuclear calculations. 
The difference between the results becomes larger with larger values of 
the kinematical variables. The cross section obtained with the monopole 
form factor is larger for each case.
The distribution in impact parameter, purely theoretical (cannot be
checked experimentally), quickly drops with growing impact parameter.
The distribution in invariant mass seems rather interesting.
While at low invariant masses the box contribution wins, at invariant
masses $M_{\gamma \gamma} >$ 30 GeV the VDM-Regge contribution is
bigger. Can we thus observe experimentally the VDM-Regge
contribution? The matter is a bit more complicated as will be explained below.
The distribution in diphoton rapidity is a bit academic and in fact 
confusing for the VDM-Regge contribution and may wrongly suggest 
that all photons are produced at midrapidities. 
We shall discuss this in detail in the following. 

%--------------------------------------------------------
%--------------------------------------------------------

%--------------------------------------------------------
%\begin{figure}[!h]  
%\center
%\includegraphics[width = 0.45\textwidth]{dsig_dW_box_VDM.eps}
%\includegraphics[width = 0.45\textwidth]{dsig_dY_box_VDM.eps}
%  \caption{\label{fig:1}
%  \small  
%Two-photon invariant mass distribution (left panel) and
%two-photon rapidity distribution (right panel) for  
%the PbPb$\to$PbPb$\gamma\gamma$ reaction
%in UPC of heavy ions.
%The distributions with realistic charge density and
%with full range of $p_{t,\gamma}$ are depicted.
%The solid line shows the results for the case
%when light-by-light scattering contributions
%come from the lepton, quark and W-boson loops.
%The dashed lines show the results when
%the sub-process is calculated in pQCD approach (VDM-Regge). 
%}
%\end{figure}
%--------------------------------------------------------

\begin{figure}[!h]  
\center
\includegraphics[scale=0.35]{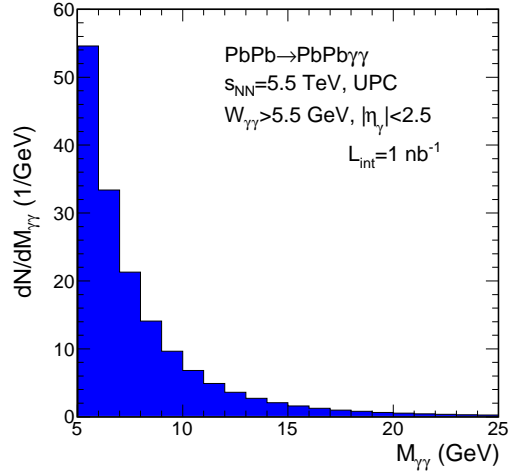}
  \caption{\label{fig:number_of_counts}
  \small
Distribution of expected number of counts in $1$~GeV bins for cuts specified in the
figure legend.  This figure should be compared with a similar figure
in \cite{d'Enterria:2013yra}. 
%Comparison of the distribution $d\sigma/dM_{\gamma\gamma}$.
%Upper panel - Fig is coped from \cite{d'Enterria:2013yra}.
%Lower fig is from our calculations.
}
\end{figure}

Can something be measured with the help of LHC detectors?
In Fig.~\ref{fig:number_of_counts} we show numbers of counts
in the $1$ GeV intervals expected for assumed integrated luminosity of $1$~nb$^{-1}$,
where in addition to the lower cut on photon-photon energy
we have imposed cuts on (pseudo)rapidities of both photons.
It looks that one can get (measure) invariant mass distribution up to 
$M_{\gamma \gamma} \approx 15$ GeV. This is much more than predicted
previously in Ref.~\cite{d'Enterria:2013yra}. We do not have clear explanation
of the difference.

%-------------------------------------------------------------------
%-------------------------------------------------------------------

%--------------------------------------------------------
\begin{figure}[!h]  
\center
\includegraphics[scale=0.35]{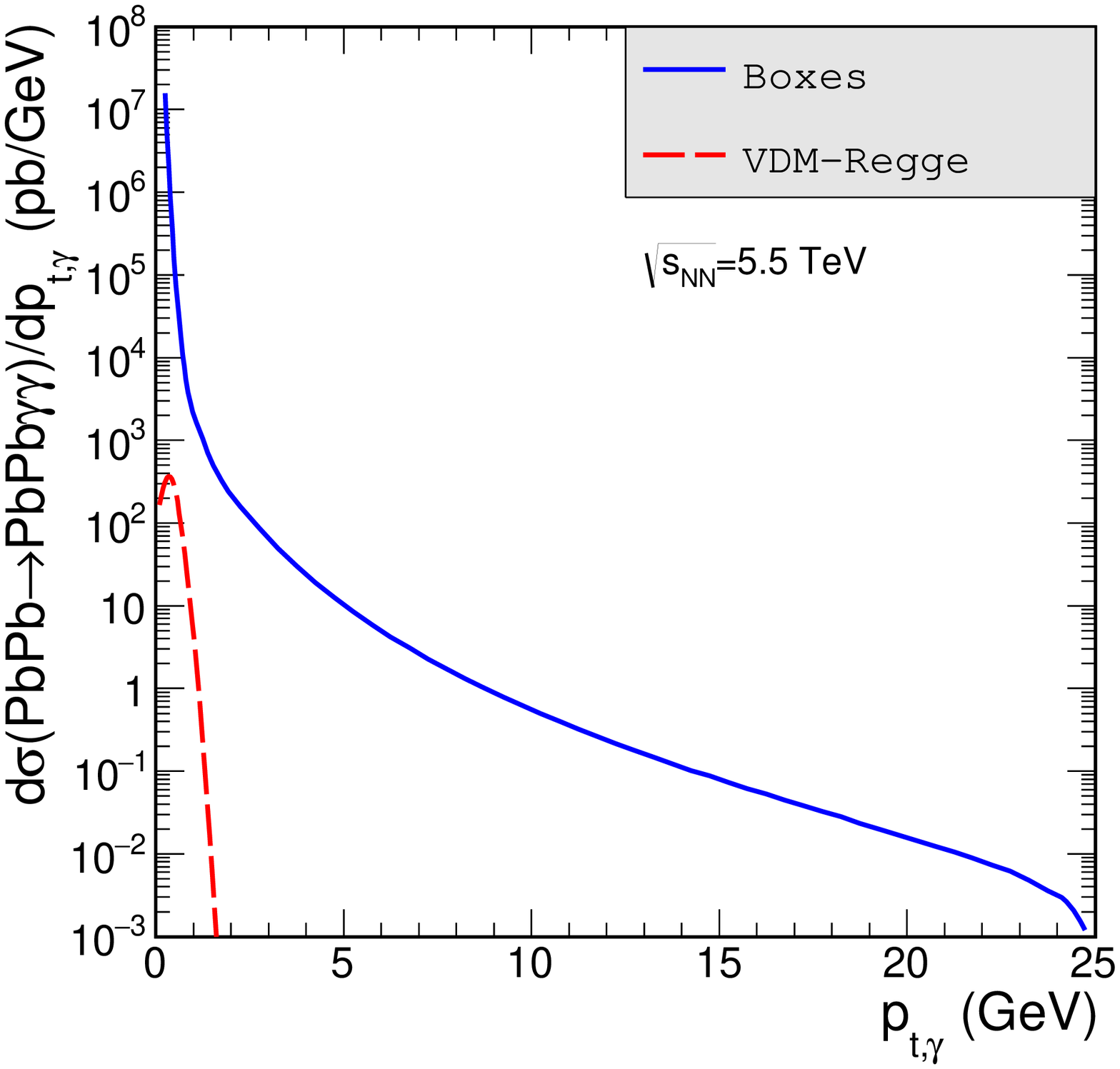}
\includegraphics[scale=0.35]{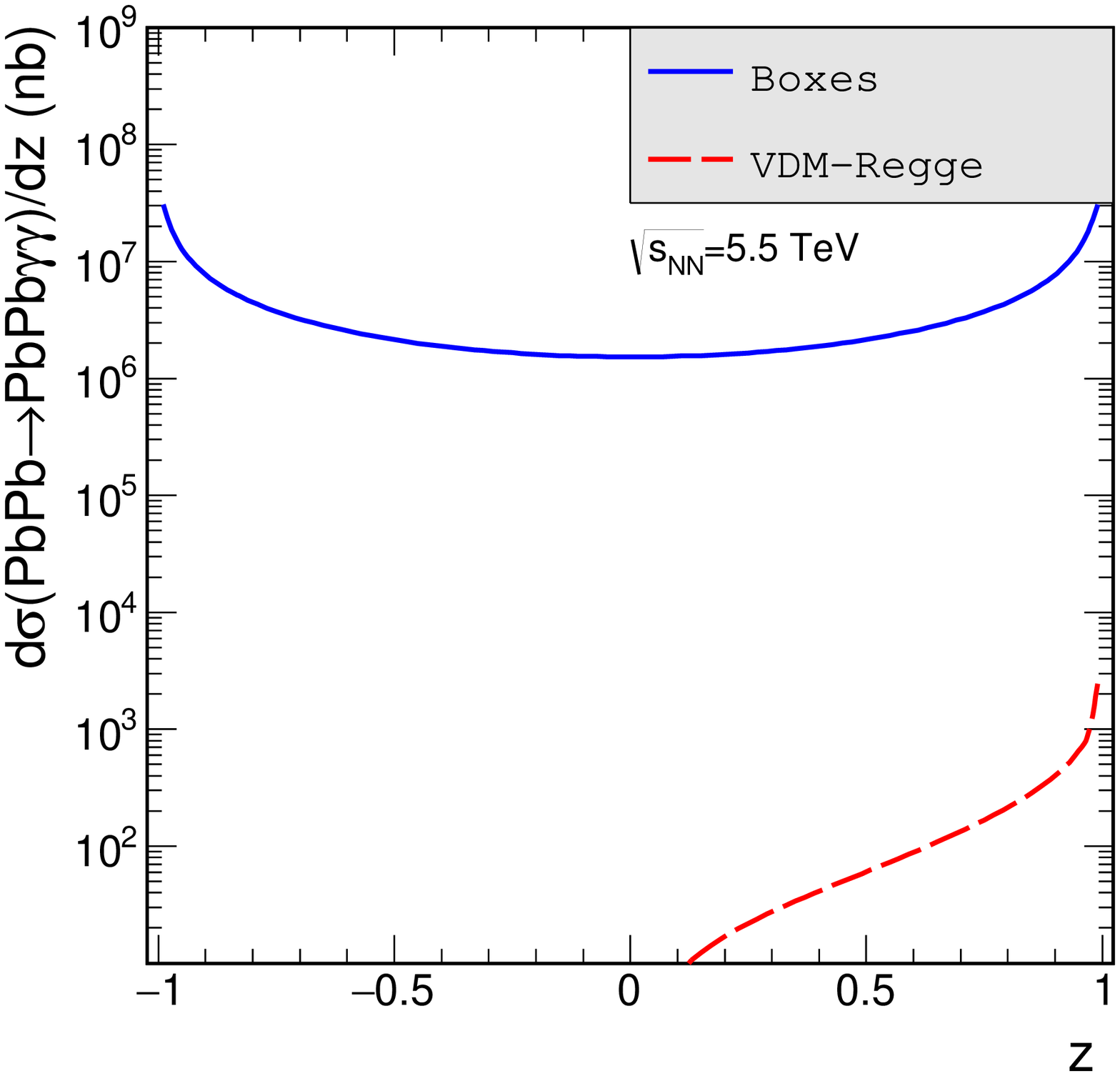}
  \caption{\label{fig:dsig_dpt_dz}
  \small
Results for the PbPb$\to$PbPb$\gamma\gamma$ reaction
in UPC of heavy ions.   
Differential nuclear cross section as a function of photon transverse momentum
$p_{t,\gamma}$ and cosine of the angle between outgoing photons $z=\cos\theta^*$
at $\sqrt{s_{NN}}=5.5$ TeV with minimal cut on 
$M_{\gamma\gamma}>1$ GeV for the VDM-Regge mechanism only.
The solid lines show the results for the case
when only box contributions (fermionic loops) are included.
The dashed lines show the results for the VDM-Regge approach only.
}
\end{figure}
%--------------------------------------------------------

Now we wish to show some selected results with essentially 
no cuts except of a minimal cut to assure that the VDM-Regge model applies. 
Fig.~\ref{fig:dsig_dpt_dz} shows differential cross section
as a function of photon transverse momentum
$p_{t,\gamma}$ (left panel) and cosine of the angle 
between outgoing photons $z=\cos\theta^*$
(right panel). The calculations are done at the LHC energy $\sqrt{s_{NN}}=5.5$ TeV.
Here we impose no cuts on kinematical variables for box contribution
and VDM-Regge mechanism except $M_{\gamma\gamma}>1$ GeV condition. 
We know that the VDM-Regge mechanism does not apply below this value. 
One can observe that the nuclear $d\sigma/dp_{t,\gamma}$ distribution 
falls very quickly for both mechanisms and is very narrow  
for the VDM-Regge contribution. 
The distribution in $z$ (right panel) shows
that without cuts on kinematical variables 
the maximal cross section occurs at $z \approx \pm 1$. 
We show only one half of the $z$ distribution for the VDM-Regge approach,
because we include only contribution from the $t$ channel in our calculation.
Contribution from the $u$ channel should have similar shape of distribution
but for the second half of the $z$ distribution 
and would be symmetric to the $t$-channel contribution around the $z=0$.
We see that already for $W_{\gamma\gamma}=M_{\gamma\gamma}>1$ GeV
it is not necessary to symmetrize the $t$ and $u$ diagrams. 

%-------------------------------------------------------------------
\begin{figure}[!h]
\includegraphics[scale=0.4]{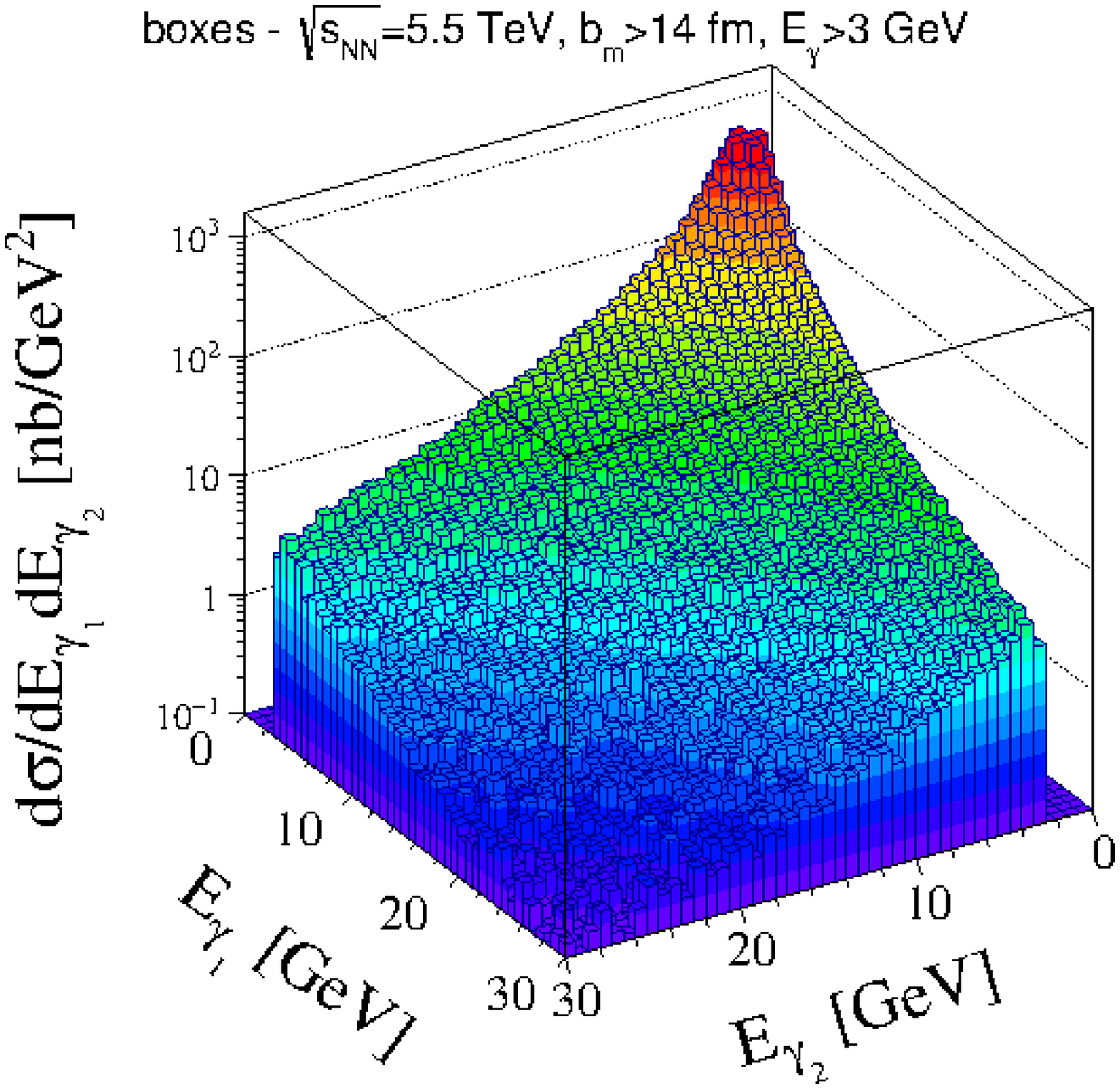}
\includegraphics[scale=0.4]{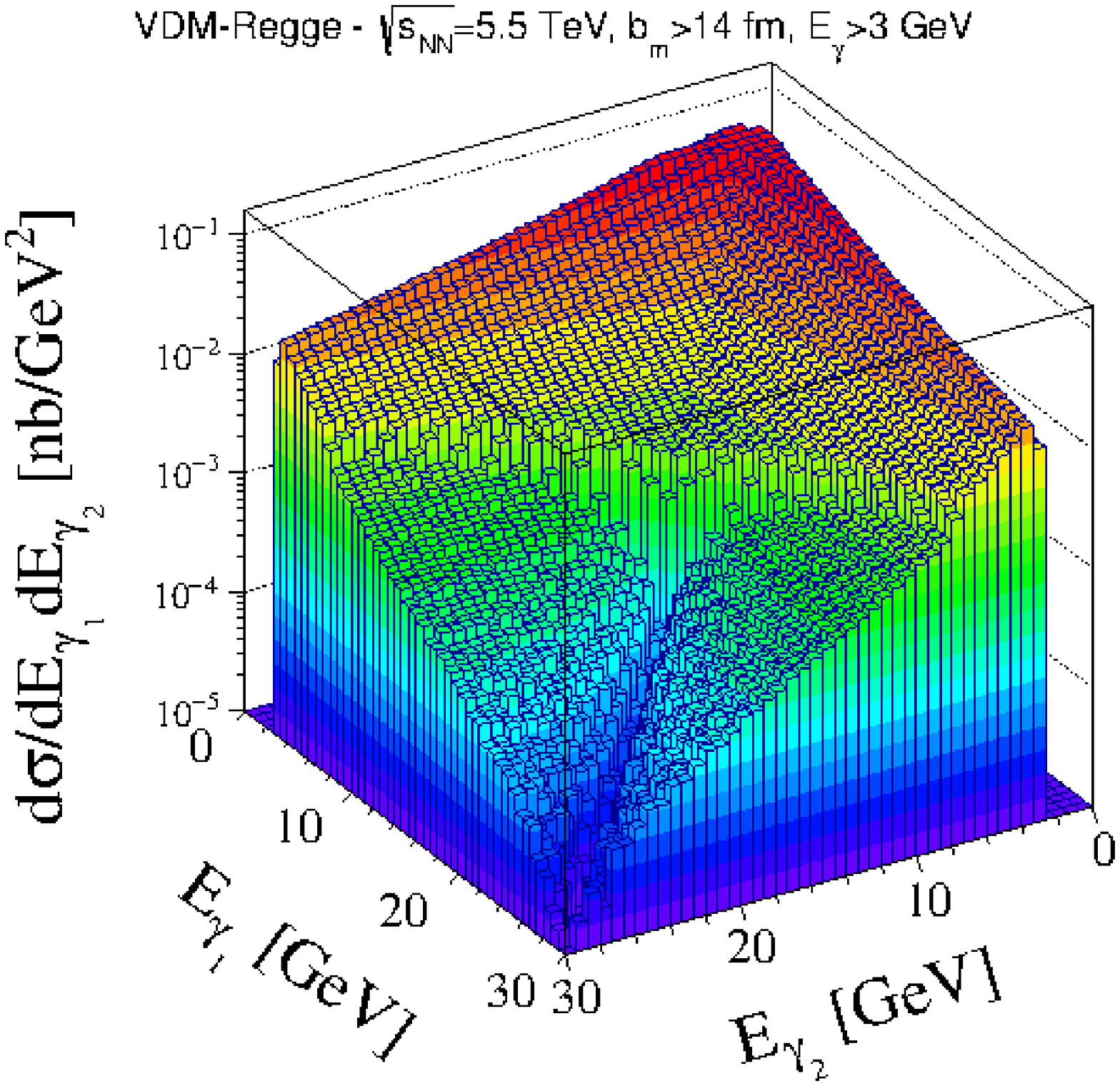}
\caption{\label{fig:dsig_dE1dE2}
\small
Two-dimensional distribution in energies of the two photons in
the laboratory frame for box (left panel) and VDM-Regge (right panel) 
contributions.
}
\end{figure}
%--------------------------------------------------------------------

The cuts on photon-photon energies are in principle not necessary.
What are in fact energies of photons in the laboratory frame of
reference? In Fig.~\ref{fig:dsig_dE1dE2} we show distribution of energies
of both photons, separately for the two mechanisms: boxes (left panel)
and VDM-Regge (right panel). In this calculations we do not impose cuts
on $W_{\gamma \gamma}$ but only minimal cuts on energies of individual
photons ($E_{\gamma} > 3$ GeV) in the laboratory frame.
Slightly different distributions are obtained for boxes and VDM-Regge
mechanisms. For the box mechanism we can observe a pronounced maximum
when both energies are small.
For both mechanisms the maximum of the cross section
occurs for rather asymmetric configurations: $E_1 \gg E_2$ or 
$E_1 \ll E_2$.

%-------------------------------------------------------------------
\begin{figure}
\includegraphics[scale=0.4]{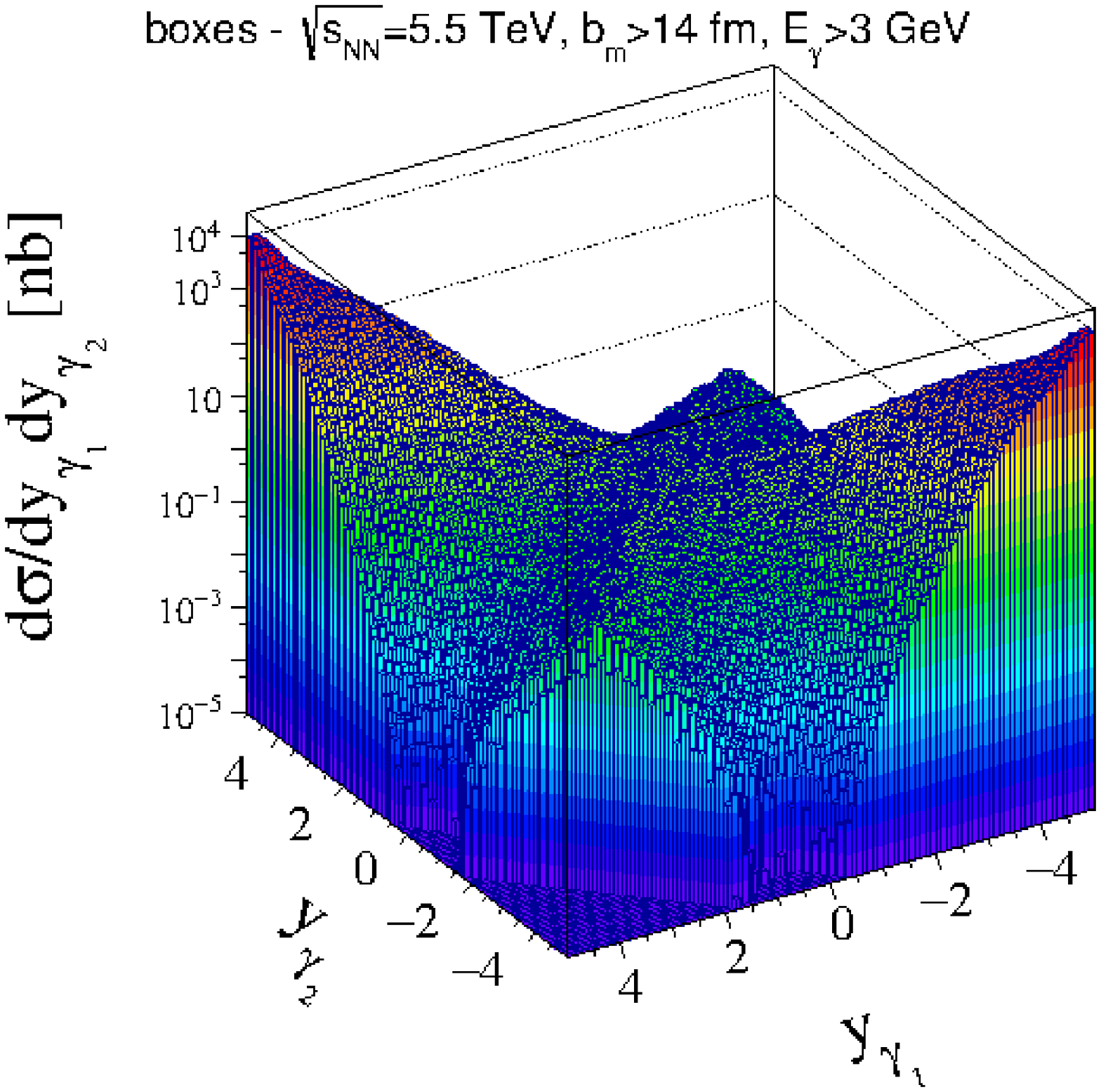}
\includegraphics[scale=0.4]{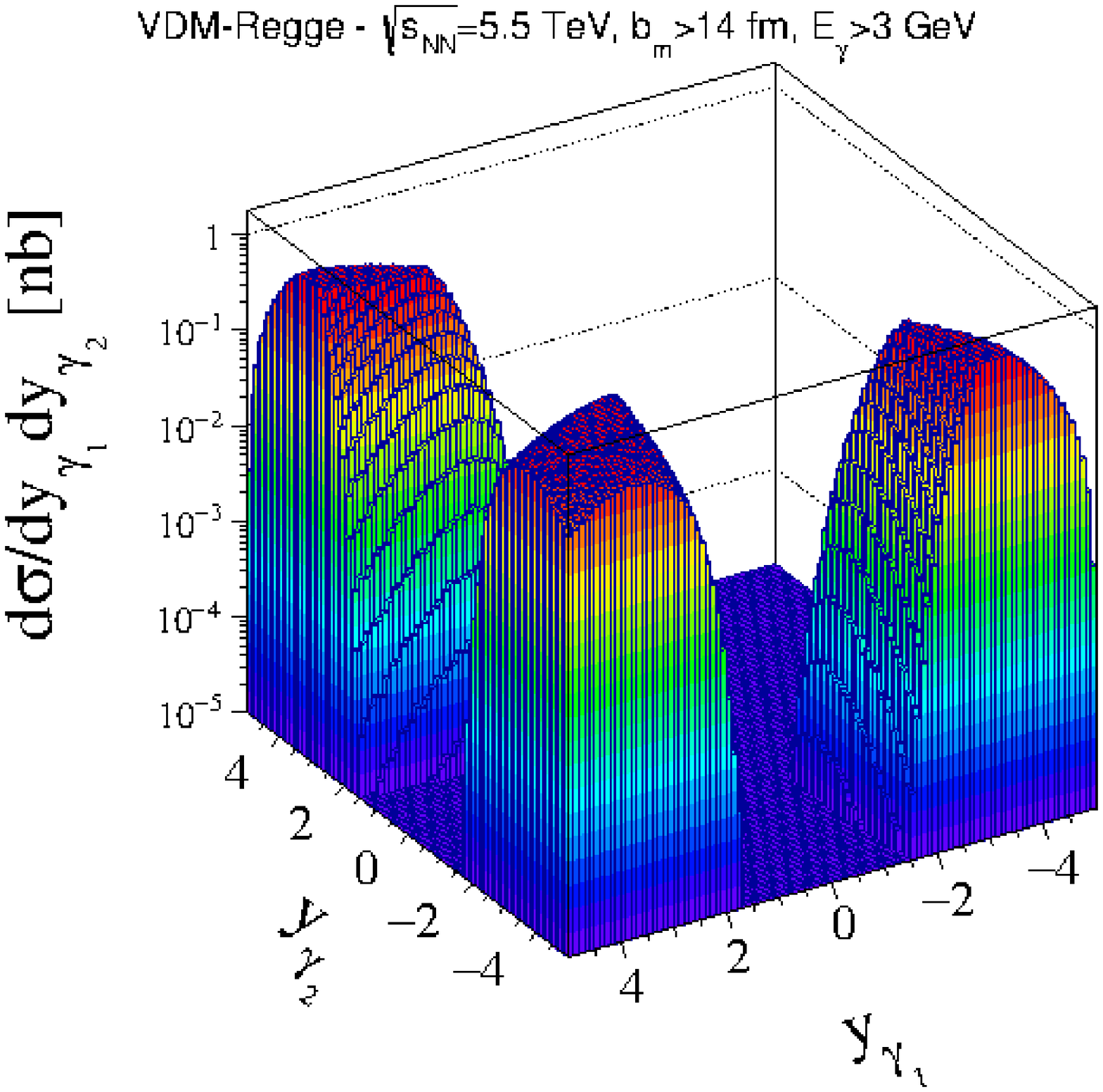}
\caption{\label{fig:dsig_dy1dy2}
\small
Two-dimensional distribution in rapidities of the two photons in
the laboratory frame for box (left panel) and VDM-Regge (right panel) 
contributions.
}
\end{figure}
%--------------------------------------------------------------------
%--------------------------------------------------------
\begin{figure}[!h]  
\center
\includegraphics[scale=0.35]{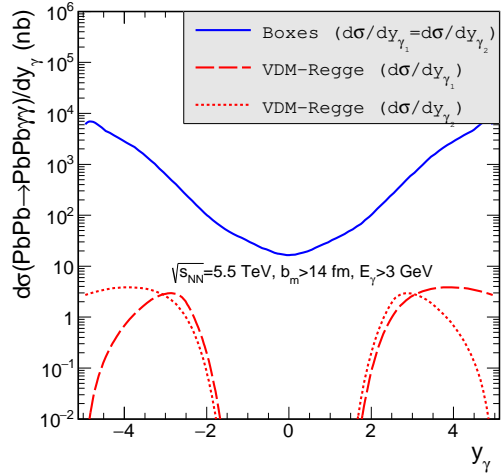}
  \caption{\label{fig:dsig_dyi_E3}
  \small
Projection on rapidity of one of the photons.
The cuts here are the same as in Fig.~\ref{fig:dsig_dy1dy2}.
For the VDM contribution, the dashed and dotted lines are projections
on the $y_{\gamma_1}$ and $y_{\gamma_2}$ axes, respectively.
}
\end{figure}
%--------------------------------------------------------

Two-dimensional distributions in photon rapidities are very different
for the box and VDM-Regge contributions (Fig.~\ref{fig:dsig_dy1dy2}). 
In the case of the VDM-Regge
contribution we observe as if non continues behaviour 
(dip in the cross section, better visible in Fig.~\ref{fig:dsig_dyi_E3}
where we show projections on both axes) which is
caused by the strong transverse momentum dependence of the elementary cross section
(see Fig.~\ref{fig:dsig_dpt_W})
which causes that some regions in the two-dimensional space are 
almost not populated.

\begin{figure}
\includegraphics[scale=0.4]{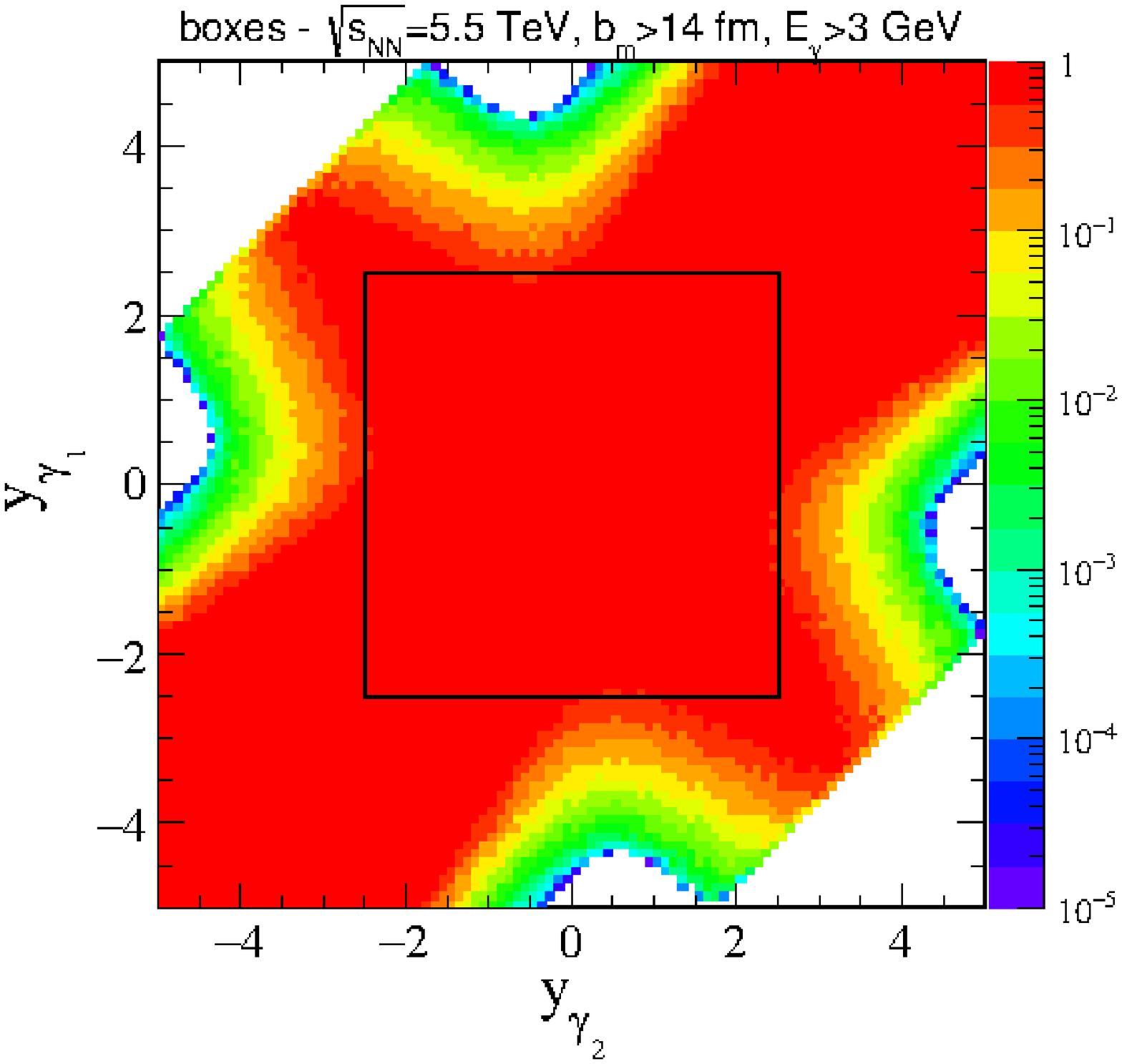}
\includegraphics[scale=0.4]{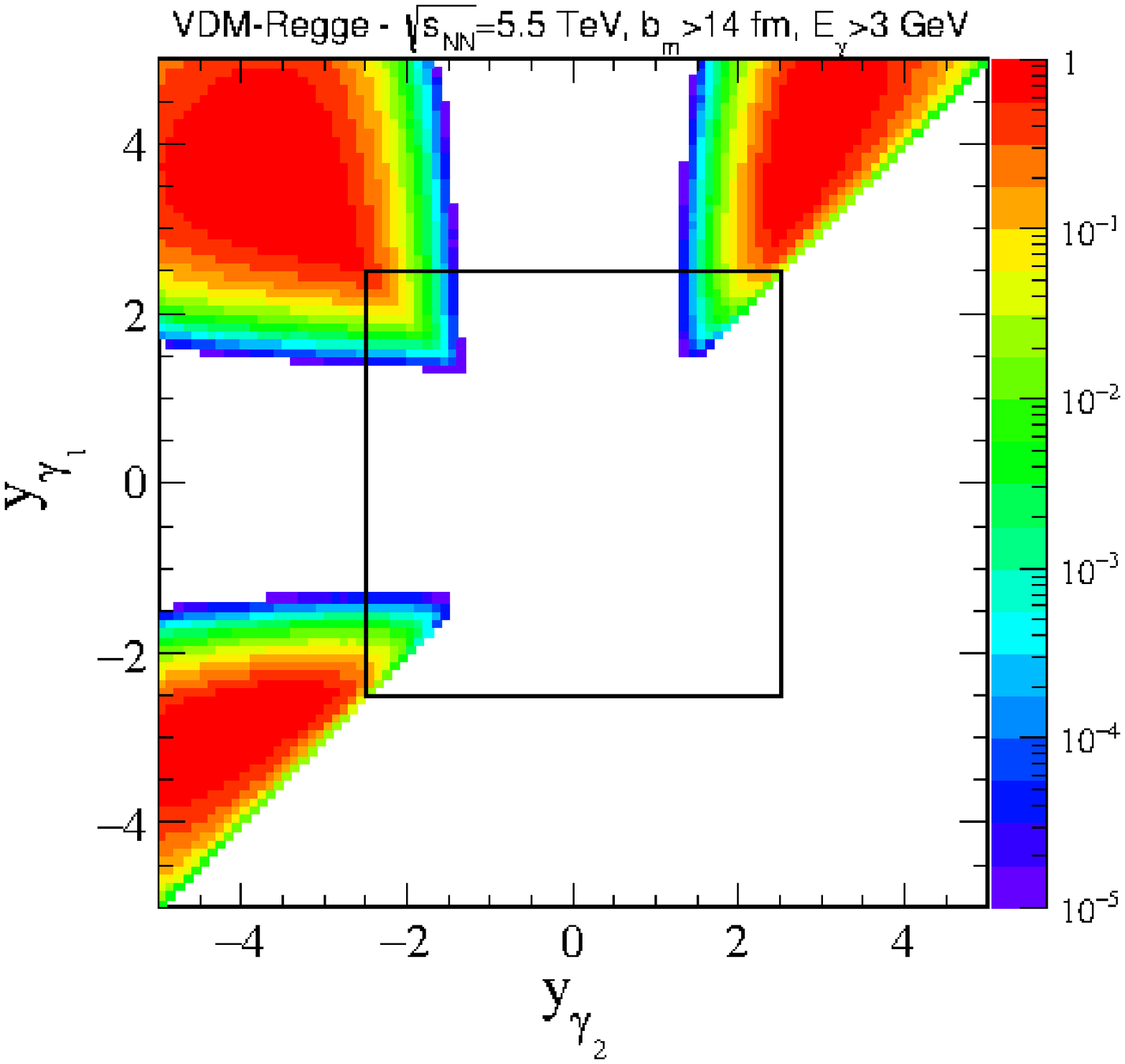}
\caption{\label{fig:dsig_dy1dy2_contour}
\small
Contour representation of two-dimensional distribution in rapidities 
of the two photons in the laboratory frame for box (left panel) 
and VDM-Regge (right panel) contributions with shown experimental rapidity
coverage of the main the ATLAS or CMS detectors.
Only one half of the ($y_{\gamma_1},y_{\gamma_2}$) space is shown for
the VDM-Regge contribution. The second half can be obtained from the symmetry
around the $y_{\gamma_1}=y_{\gamma_2}$ line.}
\end{figure}

In Fig.~\ref{fig:dsig_dy1dy2_contour} we show the same distributions
in the contour representation with the experimental limitations 
($y_{\gamma_1},y_{\gamma_2} \in (-2.5,2.5)$)
of the main detectors. For the case of the VDM-Regge contribution
we show distribution for only one half of the $(y_{\gamma_1},y_{\gamma_2})$ space. 
Clearly the VDM-Regge contribution does not fit
to the main detector and extends towards large rapidities.
Could photons originating from this mechanism be measured
with the help of so-called zero-degree calorimeters (ZDCs) associated with the ATLAS
or CMS main detectors?

%-------------------------------------------------------------------
%--------------------------------------------------------------------

%-------------------------------------------------------------------
\begin{figure}[!h]
\includegraphics[scale=0.4]{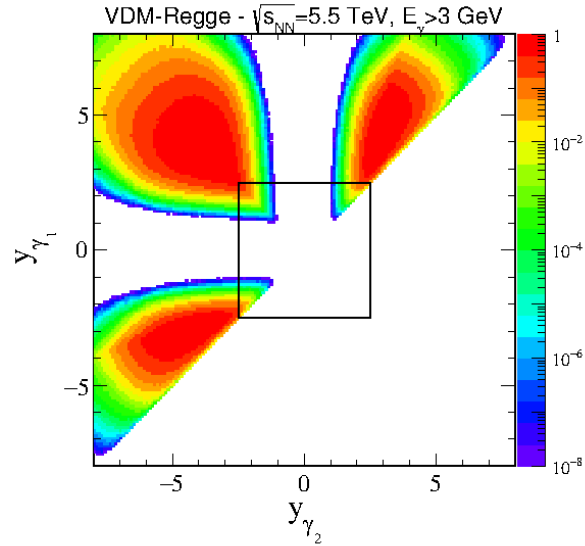}
\caption{\label{fig:dsig_dy1dy2_contour_extanded}
\small
Contour representation of two-dimensional distribution in rapidities 
of the two photons in the laboratory frame for  
the VDM-Regge contribution in the extended range of rapidities.
Only one half of the ($y_{\gamma_1},y_{\gamma_2}$) space is shown
explicitly. The second half can be obtained from the symmetry
around the $y_{\gamma_1}=y_{\gamma_2}$ line.
}
\end{figure}
%--------------------------------------------------------------------

In Fig.~\ref{fig:dsig_dy1dy2_contour_extanded} we show the VDM-Regge contribution
in much broader range of rapidity. Now we discover that maxima
of the cross section associated with this mechanism are at
$|y_{\gamma_1}|,|y_{\gamma_2}| \approx$ 5. Unfortunately this is below the limitations 
of the ZDCs $|\eta| > 8.3$ for ATLAS (\cite{ATLAS:2007aa})
or $8.5$ for CMS (\cite{Grachov:2008qg}).

%--------------------------
\section{Conclusions}
%--------------------------

We have performed detailed feasibility studies of elastic photon-photon
scattering in ultraperipheral heavy ion collisions at the LHC.
The calculation was performed in equivalent photon approximation
in the impact parameter space. This method allows to remove those
cases when nuclei collide and therefore break apart. Such cases
are difficult in interpretation and were omitted here.

The cross section for elementary photon-photon scattering has been 
calculated including box diagrams with elementary Standard Model
particles as well as a new component called here ''VDM-Regge'' for brevity. 
This soft component is based on the idea of hadronic fluctuation of 
the photon(s). The photons interact when they are in their hadronic 
(virtual vector meson) states. The standard soft Regge phenomenological 
type of interaction is used for the hadron-hadron interaction.
The VDM-Regge mechanism gives in general much smaller cross section but
for $W_{\gamma \gamma} >$ 30 GeV starts to dominate over the box
contributions, at least in the full phase space.

Several distributions in the ''standard'' EPA were calculated.
The results were compared to results of earlier calculation in the literature.
We have found cross sections by a factor of about 8 bigger
than previously published in the literature. 
We have made an estimate of the counting
rate with expected integrated luminosity. We expect some counts 
($ N > $ 1) for $W_{\gamma \gamma} = M_{\gamma \gamma} <$ 15-20 GeV.  

We have performed a detailed calculation including also distributions of
individual outgoing photons by extending the standard EPA.
We have made estimation of the integrated cross section for 
different experimental situations relevant for the ALICE or CMS experiments
as well as shown several differential distributions.
We have studied whether the VDM-Regge component,
not discussed so far, could be identified experimentally 
and have shown that it will be probably very difficult.

We have found that, very different than for the box contribution,
the VDM-Regge contribution reaches a maximum of the cross section 
when ($y_{\gamma_1} \approx 5$, $y_{\gamma_2} \approx -5$)
or ($y_{\gamma_1} \approx -5$, $y_{\gamma_2} \approx 5$). This is a rather difficult region
which cannot be studied e.g with ZDC's installed at the LHC.

So far we have studied only diphoton continuum.
The resonance mechanism could be also included in the future.
In the present studies we have concentrated on the signal.
Future studies should include also estimation of the background.
The dominant background may be expected from the $A A \to A A e^+ e^-$
when both electrons are misidentified as photons. 

\vspace{1.5cm}

{\bf Acknowledments}

This work was partially supported by the Polish grant 
No. DEC-2014/15/B/ST2/02528 (OPUS)
as well as by the Centre for Innovation and Transfer of Natural Sciences
and Engineering Knowledge in Rzesz\'ow.
A~part of the calculations within this analysis was carried out 
with the help of the cloud computer system 
(Cracow Cloud One\footnote{cc1.ifj.edu.pl}) 
of the Institute of Nuclear Physics PAN.

%------------------------------------------------------------------
%%% bibliography for thesis.
\nocite{}

{
\begin{small}
\bibliography{refs}
\end{small}
}
%------------------------------------------------------------------

\end{document}